\documentclass[manuscript]{aastex61}

\shorttitle{Poro et al.}
\usepackage{lineno}
\usepackage{amsmath}
\usepackage{graphicx}
\usepackage{xcolor}
\usepackage{longtable}
\usepackage{times}
\usepackage{tablefootnote}
\usepackage{appendix}
\usepackage{lineno}

\begin{document}

\title{The Photometric Study of Six W UMa Systems and Investigation of the Mass-Radius Relations for Contact Binary Stars}

\correspondingauthor{Atila Poro}
\email{poroatila@gmail.com, astronomydep@raderonlab.ca}

\author{Atila Poro}
\affil{Astronomy Department of the Raderon Lab., BC, Burnaby, Canada}
\affil{Binary Systems of South and North (BSN) Project, Contact Systems Department, Global Online Project}

\author{Ehsan Paki}
\affil{Binary Systems of South and North (BSN) Project, Contact Systems Department, Global Online Project}

\author{Mark G. Blackford}
\affil{Binary Systems of South and North (BSN) Project, Contact Systems Department, Global Online Project}
\affil{Variable Stars South (VSS), Congarinni Observatory, Congarinni, NSW, 2447, Australia}

\author{Fatemeh Davoudi}
\affil{Astronomy Department of the Raderon Lab., BC, Burnaby, Canada}
\affil{Binary Systems of South and North (BSN) Project, Contact Systems Department, Global Online Project}

\author{Yasemin Aladag}
\affil{Space Science and Solar Energy Research and Application Center (UZAYMER), University of Çukurova, 01330, Adana, Turkey}

\author{Shiva Zamanpour}
\affil{Binary Systems of South and North (BSN) Project, Contact Systems Department, Global Online Project}

\author{Soroush Sarabi}
\affil{AI Department of the Raderon Lab., BC, Burnaby, Canada}

\author{Afshin Halavati}
\affil{Binary Systems of South and North (BSN) Project, Contact Systems Department, Global Online Project}
\affil{Bkaran Observatory, Kerman Province, Kerman, Iran}

\author{Nazim Aksaker}
\affil{Space Science and Solar Energy Research and Application Center (UZAYMER), University of Çukurova, 01330, Adana, Turkey}
\affil{Adana Organised Industrial Zones Vocational School of Technical Science, University of Çukurova, 01410, Adana, Turkey}

\author{Halil Bagis}
\affil{Space Science and Solar Energy Research and Application Center (UZAYMER), University of Çukurova, 01330, Adana, Turkey}

\author{Jabar Rahimi}
\affil{Binary Systems of South and North (BSN) Project, Contact Systems Department, Global Online Project}

\author{Hamidreza Guilani}
\affil{Binary Systems of South and North (BSN) Project, Contact Systems Department, Global Online Project}

\author{Aysun Akyuz}
\affil{Space Science and Solar Energy Research and Application Center (UZAYMER), University of Çukurova, 01330, Adana, Turkey}
\affil{Department of Physics, University of Çukurova, 01330, Adana, Turkey}

\author{Faezeh Jahediparizi}
\affil{Binary Systems of South and North (BSN) Project, Contact Systems Department, Global Online Project}
\affil{Department of Physics, Shahid Bahonar University of Kerman, Kerman, Iran}

\author{Ozge Doner}
\affil{Space Science and Solar Energy Research and Application Center (UZAYMER), University of Çukurova, 01330, Adana, Turkey}

\author{Zohreh Ashrafzadeh}
\affil{Binary Systems of South and North (BSN) Project, Contact Systems Department, Global Online Project}

\begin{abstract}
We present the photometric analysis of six short-period systems (EI Oct, V336 TrA, NX Boo, V356 Boo, PS Boo, and V2282 Cyg). This is the first photometric analysis of these systems except for V336 TrA. Observations were conducted for 27 nights at three observatories in the northern and southern hemispheres. We calculated a new ephemeris for each of the systems using our minimum times and additional literature. The Markov Chain Monte Carlo (MCMC) approach was used to determine the eclipse timing variation trends of the systems. We found a likely orbital growth for V336 TrA and PS Boo; four other systems show a linear trend in orbital period changes, which is most likely due to the accumulation of measurement errors in their linear ephemeris parameters. The light curve analysis was performed using the Physics of Eclipsing Binaries (PHOEBE) 2.3.59 version code with the MCMC approach. The absolute parameters of the systems were calculated by using the Gaia Early Data Release 3 (EDR3) parallax. The positions of the systems were also depicted on the Hertzsprung-Russell (H-R) and $logJ_0-logM$ diagrams. According to a sample, we were able to present relations for the mass-radius ($M-R$) relationships of contact binary systems. There is also a strong relationship between the mass ratio and the radius ratio in the W UMa systems for which we also provided a new relation. We compared the $M-R$ updated relationships in this study with seven systems in other studies obtained using the spectroscopic method. In addition, we estimated some of the absolute parameters for 1734 EW systems, based on the new relationships.
\end{abstract}


\section{Introduction}\label{sect1}
W UMa stars are low-mass eclipsing Contact Binary systems (CBs) that are the most common in the cosmos among eclipsing binaries (\citealt{1948HarMo...7..249s}). The orbital period of W UMa-type CBs (EWs) is less than one day. EWs are divided into two subtypes: A and W (\citealt{1970VA.....12..217B}). A-subtype systems are early spectral types with higher mass and luminosity than W-subtype systems. In A-subtype systems the mass ratio ($q$) is usually less than 0.5 suggesting modest or minor activity. In W-subtype systems the less massive component is hotter and the orbital period changes continuously throughout time (\citealt{1970VA.....12..217B}).
Most EWs show continuous light variations and have asymmetrical light curves with brightness differences at the phases of 0.25 and 0.75. The existence of cold or hot spots on their surface is a sign of a phenomenon called the O'Connell effect (\cite{1951PRCO....2...85O}). The magnitude of this imbalance in light can fluctuate over time due to the development and movement of spots on the stellar surface.
The components of EWs fill their Roche lobes so that the strongly distorted stars touch each other at the inner Lagrangian point. These components are so close together that their structure indicates the interaction and the transfer of energy and mass between them. The formation, the evolution, and the ultimate destiny of EWs are currently unknown (\citealt{1994ASPC...56..228B} \citealt{2003MNRAS.342.1260Q} \citealt{2005ApJ...629.1055Y} \citealt{2012JASS...29..145E}). Hence, modeling a variety of them by using extensive data is a powerful method for accurately determining the fundamental physical parameters of stars.
The precise determination of the value of $q$ has a significant impact on the values of the star masses, radii, and luminosities derived by modeling the observable data of EWs. As the spectral lines of the EWs are widened and mixed, the accurate identification of their $q$ value is impossible to determine (\cite{2000A&A...358.1007F} \cite{2005MNRAS.357..497B}). Among the limited number of EWs that have spectral estimates of their $q$, the EWs that undergo partial eclipses have their $q$ values poorly determined by the photometric method (\citealt{2001AJ....122.1007R} \citealt{2005Ap&SS.296..221T}). However, due to the well-defined $q$ value of the EWs that experience total eclipses, their light curve solutions are the best option for calculating the global stellar parameters.

In this study, new ephemerides were calculated for the six EW systems, and we present photometric light curve analysis of four of them. Four systems from the Northern Hemisphere (NX Boo, PS Boo, V356 Boo, and V2282 Cyg) and two systems from the Southern Hemisphere (EI Oct and V336 TrA) have been selected. The following is how the paper is structured: The information about photometric observations and a data reduction method is given in section 2. Extracting minimum times and the new ephemeris of each system is presented in section 3. Section 4 discusses the photometric light curve solutions for these systems. The methods used to determine physical parameters are described in section 5. Discussion of some results, and the mass and radius relationships for W UMa systems are presented in sections 6 and 7, respectively. Finally, the summary and conclusion are provided in section 8.

\vspace{1.5cm}
\section{Observation and data reduction}\label{sec:obs}
The observations were performed at three observatories located in the northern and southern hemispheres. Over the course of 27 nights in 2020-2021, observations were made using the photometric method. Table \ref{tab1} contains information on the observations made for each of the systems. The name of the system appears in the first column, in the second column, the observatory that observed each system, in the third column, the year/month of observation, in the fourth column, the number of observation nights for each system, in the fifth column, the filters used for observation, and in the sixth and seventh columns, the names of comparison and check stars, respectively. The general features of the systems are shown in table \ref{tab2}.
Congarinni, a southern hemisphere observatory, is located in Australia at 152° 51’ 38” East, 30° 44’ 04” South, at an altitude of 20 meters above mean sea level. This observatory uses a GSO 14-inch Ritchey Chretien telescope and an SBIG STT3200-ME CCD with Astrodon Johnson-Cousins $BVI$ filters for photometry.
Two observatories in the northern hemisphere also conducted observations. Bkaran Observatory is located in Iran at long. 57° 01’ 13’’ East, lat. 30° 16’ 55’’ North, at an altitude of 1764 meters. A 10-inch Schmidt-Cassegrain telescope and a Nikon D5300 Digital Single-Lens Reflex (DSLR) camera with a $V$ Standard Johnson filter are used at this observatory. Another site is located at Adana's Çukurova University. The UZAYMER observatory is located at 35° 21’ 19’’ East, 37° 03’ 35’’ North, at an altitude of 130 meters. A 50cm Ritchey Chretien telescope, an Apogee Aspen CG type CCD, and $BVR$ standard Johnson filters are used at this observatory.

We did all image processing and plotting of raw images with MaxIm DL software (\citealt{2000IAPPP..79....2G}) and AstroImageJ (AIJ) software package (\citealt{2017AJ....153...77C}). Calibration of images (e.g. bias, dark, and flat-fielding) was performed on all observation nights. The AIJ software package was used to detrend the effect of airmass on all data.

\begin{table*}
\caption{Specifications of observations performed for six systems, all of which were observed during 2020-2021.} 
\centering
\begin{center}
\footnotesize
\begin{tabular}{c c c c c c c}
 \hline
 \hline
        System & Observatory & Year/Month & Nights & Filter & Comparison star & Check Star \\
	   \hline
       EI Oct	& Congarinni	& 2020/8	& 6	& $BV$	& GSC 9516-1312	& GSC 9516-1389 \\ 
       V336 TrA	& Congarinni	& 2020/5,6 - 2021/4,6	& 8	& $BVI$	& UCAC4 134-144041	& GSC 9027-4852 \\ 
       NX Boo	& UZAYMER	& 2020/7	& 5	& $VR$	& TYC 2565-474-1	& TYC 2565-303-1 \\
       V356 Boo	& Bkaran	& 2020/6	& 2	& $V$	& GSC 912-1091	& GSC 912-759\\
       PS Boo	& UZAYMER	& 2020/6	& 3	& $BVR$	& GSC 3488-452	& GSC 3488-1161\\
       V2282 Cyg	& UZAYMER	& 2020/8	& 3	& $VR$	& TYC 3921-777-1	& TYC 3921-1966-1\\
       \hline
       \hline
\end{tabular}
\end{center}
\label{tab1}
\end{table*}

\begin{table*}
\caption{Coordinates of the systems are from the Simbad database, and magnitude ranges and orbital periods are from the American Association of Variable Star Observers (AAVSO) Variable Star Index (VSX) database.}
\centering
\begin{center}
\footnotesize
\begin{tabular}{c c c c c}
 \hline
 \hline
        System	& RA.(J2000)	& DEC.(J2000)	& Mag. range/Filter\footnote{$CV$= Clear passband; $V$ band used for comparison star; $R1$= ROTSE-I (450-1000nm); $C$= Clear passband.}	& Period(d) \\
	   \hline
       EI Oct	& 12 32 42.93651	& -87 26 22.77264	& 11.35-12.10/$V$	& 0.3385245(\citealt{2006SASS...25...47W}) \\
       V336 TrA	& 15 59 06.40432	& -63 17 49.39488	& 10.46-12.20/$V$	& 0.266768(\citealt{2004IBVS.5600...27S}) \\ 
       NX Boo	& 15 00 59.54926	& +34 11 41.03125	& 12.80-13.20/$CV$	& 0.251134(\citealt{2006SASS...25...47W}) \\
       V356 Boo	& 14 20 44.32094	& +11 21 06.92388	& 11.65-12.20/$V$	& 0.286482(\citealt{2014OEJV..162....1P}) \\
       PS Boo	& 15 19 44.15811	& +50 20 57.14178	& 12.30-12.60/$R1$	& 0.2816398(\citealt{2018MNRAS.477.3145J}) \\
       V2282 Cyg	& 19 25 37.74403	& +53 25 20.28480	& 12.02-12.30/$C$	& 0.335950(\citealt{2006SASS...25...47W}) \\
       \hline
       \hline
\end{tabular}
\end{center}
\label{tab2}
\end{table*}

\vspace{1.5cm}
\section{Orbital period variations}\label{sec:O-C}
\begin{figure}
\begin{center}
\includegraphics[width=\columnwidth]{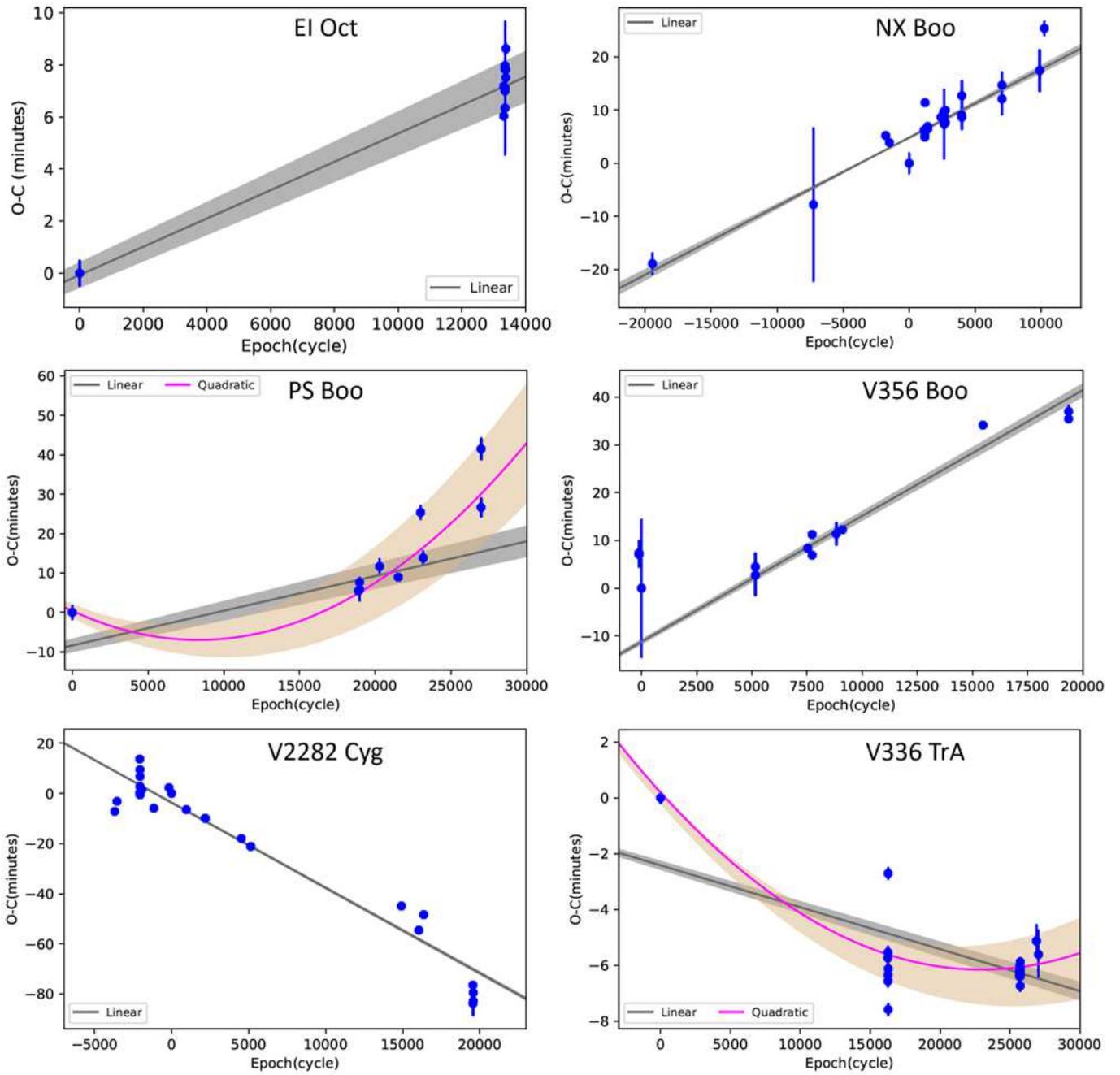}
\caption{The O-C diagram of six eclipsing binaries with linear (gray line) and quadratic (magenta curve) models. The shaded regions show the model parameters' 68th percentile values, while the curves represent their median values.}
\label{Fig1}
\end{center}
\end{figure}

The primary and secondary minimum times for each system were extracted using the observed light curves. These 37 minima times were obtained by using Gaussian and Cauchy distributions to fit the models to the light curves and existing minima. The MCMC approach was also used to calculate the amount of uncertainty (\citealt{2021AstL...47..402P}).We benefited the PyMC3 package to execute the MCMC sampling (\citealt{2016ascl.soft10016S}). Then, we fitted all minima times using a probabilistic linear fitting scheme based on sampling with the MCMC approach for extracting parameters of new ephemeris and determined a new ephemeris for each system (Table \ref{tab3}). We applied 20 walkers and 20000 iterations for each walker, with a 1000 burn-in period in the MCMC sampling. Figure \ref{Fig1} shows the O-C diagrams of the six systems. All minima collected from the previous observations along with the minima extracted from the observations of this study are documented in appendix tables \ref{A1} to \ref{A6}.

The linear model of EI Oct, NX Boo, V356 Boo, and V2282 Cyg systems emerged which could be due to the error accumulation of their light elements in their ephemeris. We applied two models for interpreting the orbital period changes of the PS Boo and V336 TrA systems. The Bayesian Information Criteria (BIC) with $\Delta BIC=16$ and $\Delta BIC=9$ values supported the quadratic models for PS Boo and V336 TrA, respectively. 
For a circular orbit, the orbital period derivative can be calculated by using the equation 1:

\begin{equation}\label{eq1}
T_{mid}(E)=T_0+PE+\frac{1}{2}\frac{dP}{dt}E^2
\end{equation}

where mid-eclipse times ($T_{mid}$) are described by $T_0$ as the reference mid-eclipse time, $P$ as the orbital period, and $E$ as the epoch of eclipses (\citealt{2017AJ....154....4P}).

We calculated the change in the orbital period derivative of V336 TrA and PS Boo by using the quadratic coefficient of models as $\frac{dP}{dt}=(6.257\pm6.252)\times10^{-11}$ day and $\frac{dP}{dt}=(5.244\pm5.227)\times10^{-10}$ day, respectively. V336 TrA and BS Poo are likely to experience orbital growth at rates of $0.001975\pm0.00091$ s/yr and $0.0165\pm0.0051$ s/yr, respectively.

\begin{figure}
\begin{center}
\includegraphics[width=\columnwidth]{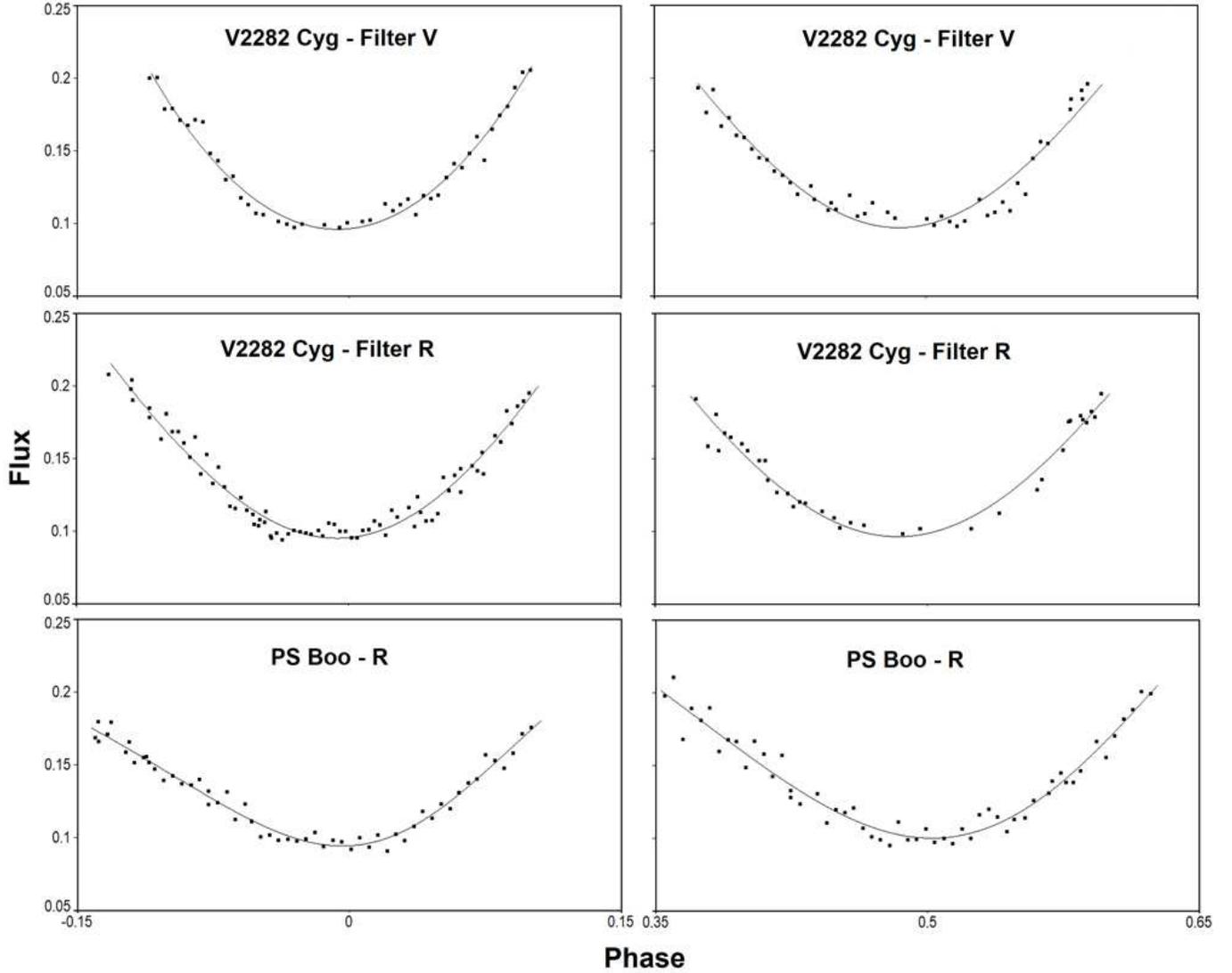}
\caption{The minima parts of the light curves for the PS Boo system in the R filter and for the V2282 Cyg system in the VR filters.}
\label{Fig2}
\end{center}
\end{figure}

The The primary and secondary minima from light curves of the two systems V2282 Cyg and PS Boo, are shown in Figure \ref{Fig2}.

\begin{table*}
\caption{New ephemeris of the systems.} 
\centering
\begin{center}
\footnotesize
\begin{tabular}{c c}
 \hline
 \hline
System & New ephemeris \\
\hline
EI Oct & $(2454556.50649_{\rm-0.00032}^{+0.00033})+(0.3385248777461_{\rm-0.0000000247102}^{+0.0000000245817})\times$$E$ \\ 
V336 TrA & $(2452151.54109_{\rm-0.00010}^{+0.00009})+(0.2667678957684_{\rm-0.0000000038659}^{+0.0000000038812})\times$$E$ \\ 
NX Boo & $(2456461.77811_{\rm-0.00007}^{+0.00007})+(0.2511348944395_{\rm-0.0000000360605}^{+0.0000000360330})\times$$E$ \\
V356 Boo & $(2453481.80993_{\rm-0.00031}^{+0.00031})+(0.2864838284412_{-0.0000000304629}^{+0.0000000305302})\times$$E$ \\
PS Boo & $(2451403.81994_{-0.00108}^{+0.00107})+(0.2816404113559_{\rm-0.0000000530058}^{+0.0000000530536})\times$$E$ \\
V2282 Cyg &	$(2452500.00619_{-0.00014}^{+0.00014})+(0.3359476395682_{\rm-0.0000000104444}^{+0.0000000104915})\times$$E$ \\
\hline
\hline
\end{tabular}
\end{center}
\label{tab3}
\end{table*}

\vspace{1.5cm}
\section{Light curve analysis}\label{Analysis}	
Light curve analysis of the EI Oct, V336 TrA, NX Boo, and V356 Boo systems has been carried out with the PHOEBE\footnote{http://phoebe-project.org} 2.3.59 version and the MCMC approach (\citealt{2020ApJS..250...34C}).

The gravity-darkening coefficients and the bolometric albedo were assumed to be $g_1=g_2=0.32$ (\citealt{1967ZA.....65...89L}) and $A_1=A_2=0.5$ (\citealt{1969AcA....19..245R}), respectively. The initial temperature for one of the components was set based on the Gaia DR2 temperature, and the final value and uncertainty were determined using the PHOEBE processing. The star selection for the initial temperature was determined according to the morphology of the light curves. Therefore, with the exception of NX Boo, for the rest of the systems, the initial temperature was set on the primary star. The \cite{2004A&A...419..725C} method was used to model the stellar atmosphere and the limb-darkening coefficients were adopted as a free parameter in the PHOEBE code.

\begin{figure*}
\begin{center}
\includegraphics[width=13cm,height=23cm]{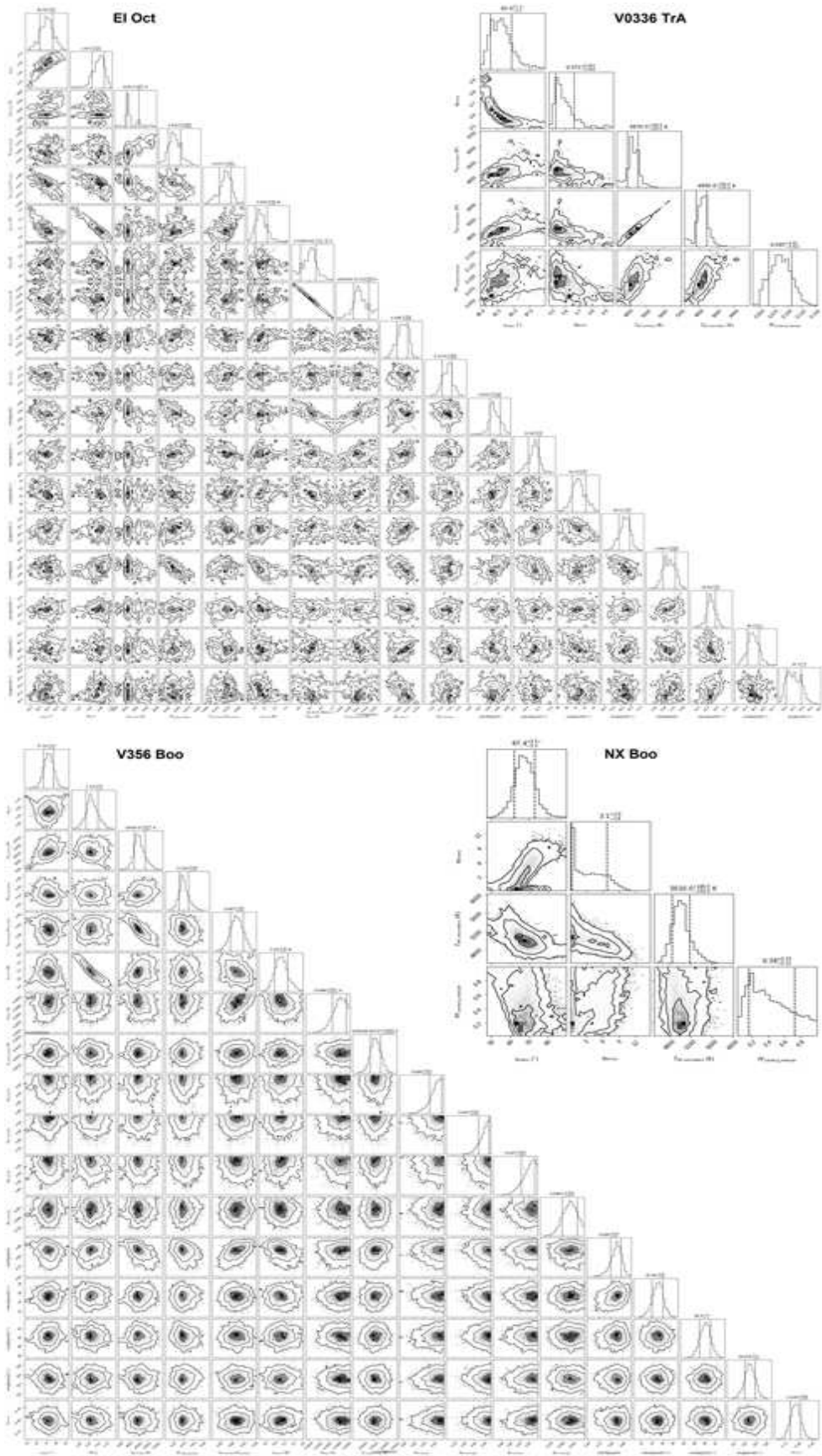}
\caption{The corner plots of the four systems were determined by MCMC modeling.}
\label{Fig3}
\end{center}
\end{figure*}

In binary systems, the mass ratio is one of the most important parameters in light curve analysis. The mass ratio of these four totally contact binary systems can be determined, however only the photometric data is available for them (\citealt{2005Ap&SS.296..221T}). First, we did a $q$-search with PHOEBE, and then we tried to fit a good synthetic light curve to the observational data. We then used PHOEBE's optimization tool to improve the output. Accordingly, initial analyses were improved using the MCMC approach and the uncertainty values were determined (Table \ref{tab4}). In the MCMC approach based on the emcee package (\citealt{2013PASP..125..306F}), we made an effort to utilize appropriate walkers and iterations. For EI Oct, V356 Boo, NX Boo, and V336 TrA, we employed 96, 96, 46, 46 walkers and 3000, 3000, 1800, and 500 iterations for each walker, respectively. The initial positions of these walkers were chosen from a Gaussian distribution based on our initial parameter estimations, the width of which was adjusted based on how sensitive each system's light curves were to different parameter values. We examined various adjusted parameters for the MCMC processing, and finally, different parameters were set for each system (\ref{Fig3}). The $q$, $T_2$, $f$, and $i$ adjusted parameters are included in all four systems.

\begin{figure}
\begin{center}
\includegraphics[width=9cm,height=20cm]{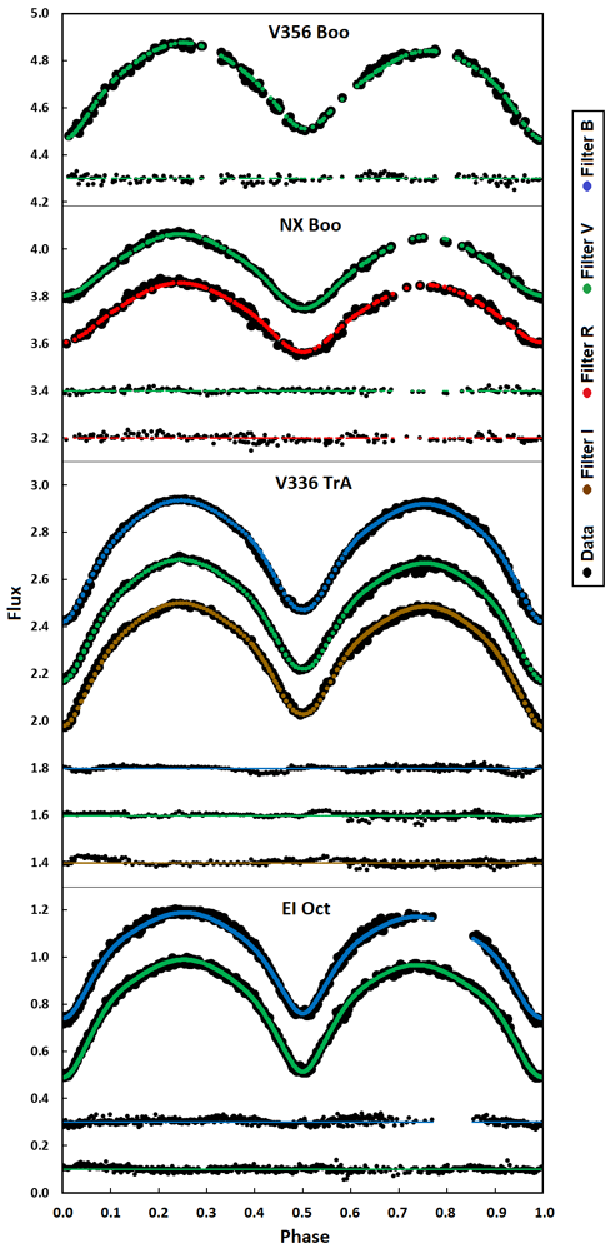}
\caption{The observed light curves of the systems (black dots) and the synthetic light curves were obtained from the light curve solutions in each filter. The orbital phase was retained but shifted the relative flux arbitrarily.}
\label{Fig4}
\end{center}
\end{figure}

Figure \ref{Fig4} shows the observed and final synthetic light curves based on this processing for each of the systems.

\begin{table*}
\caption{Photometric solutions of the systems.}
\centering
\begin{center}
\footnotesize
\begin{tabular}{c c c c c}
 \hline
 \hline
Parameter & EI Oct	& V336 TrA	& NX Boo & V356 Boo \\
\hline
$T_1$ (K) & $5330_{\rm-(44)}^{+(252)}$ & 	$4970_{\rm-(250)}^{+(250)}$ & $4692(70)$ &	$10560_{\rm-(740)}^{+(1020)}$ \\

$T_2$ (K) & $5200.5_{\rm-(45)}^{+(52)}$ &	$4900_{\rm-(270)}^{+(250)}$ & $5010_{\rm-(150)}^{+(180)}$ &	$9472_{\rm-(221)}^{+(243)}$ \\

$q$ & $1.671_{\rm-(52)}^{+(39)}$ &	$0.573_{\rm-(10)}^{\rm+(5)}$ &	$3.10_{\rm-(2.8)}^{+(3.6)}$ & $1.330_{\rm-(140)}^{+(180)}$ \\

$\Omega_1=\Omega_2$ & $4.746(175)$ &	$2.984(169)$ &	$6.513(50)$ &	$4.089(69)$ \\

$i^{\circ}$ &	$84.44_{\rm-(39)}^{+(36)}$ & $83.00_{\rm-(1.5)}^{+(1.7)}$ &	$67.40_{\rm-(5.2)}^{+(5.5)}$ &	$77.19_{\rm-(51)}^{+(51)}$ \\

$f$ & $0.055_{\rm-(8)}^{+(9)}$ &	$0.087_{\rm-(21)}^{+(20)}$ & $0.380_{\rm-(220)}^{+(35)}$ & $0.318_{\rm-(31)}^{+(44)}$ \\

$l_1/l_{tot}$ &	$0.417(2)$ & $0.644(3)$ & $0.198(4)$ &	$0.497(3)$ \\
$l_2/l_{tot}$ & $0.583(2)$ & $0.356(3)$	& $0.802(2)$	& $0.503(2)$ \\
$r_{1(mean)}$ & $0.340(29)$ & $0.436(25)$ & $0.310(8)$	& $0.384(17)$ \\
$r_{2(mean)}$ & $0.430(27)$ & $0.339(27)$ & $0.503(7)$	& $0.433(15)$ \\
Phase shift	& $0.006(1)$ & $-0.003(2)$ & $0.016(2)$	& $-0.019(1)$ \\
\hline
Spot on the primary component: & & & & \\
$Colatitude(deg)$ & $91.25_{\rm-(67)}^{+(82)}$ & $90(2)$ & & \\
$Longitude(deg)$ & $69.51_{\rm-(1.03)}^{+(87)}$ & $269(3)$ & & \\	
$Radius(deg)$ & $25.04_{\rm-(76)}^{+(52)}$ & $20(1)$ & & \\
$T_{spot}/T_{star}$ & $0.852_{\rm-(8)}^{+(4)}$ & $1.04(1)$ & & \\
Spot on the secondary component: & & & & \\
$Colatitude(deg)$ & $88.2_{\rm-(8)}^{+(1.1)}$ & & $71(2)$ & $90.0_{\rm-(1.1)}^{+(1.0)}$ \\
$Longitude(deg)$ & $91.5_{\rm-(1.3)}^{+(1.4)}$ & & $310(3)$ & $314.93_{\rm-(93)}^{+(1.0)}$ \\
$Radius(deg)$ & $24.62_{\rm-(59)}^{+(75)}$ & & $20(1)$ & $25.05_{\rm-(1.06)}^{+(96)}$ \\
$T_{spot}/T_{star}$ & $0.966_{\rm-(7)}^{+(7)}$ & & $0.90(2)$ & $0.836_{\rm-(40)}^{+(32)}$ \\
\hline
\hline
\end{tabular}
\end{center}
\label{tab4}
\end{table*}

\begin{figure*}
\begin{center}
\includegraphics[width=18cm,height=3.8cm]{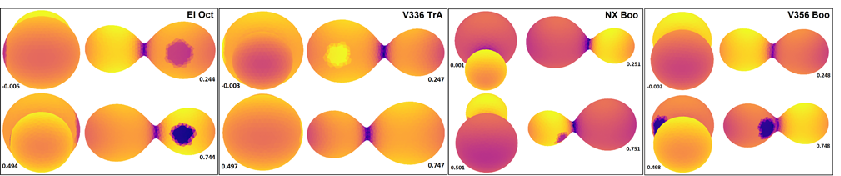}
\caption{Geometric structure of the systems based on their light curve solutions.}
\label{Fig5}
\end{center}
\end{figure*}

Asymmetry in the brightness of maxima in the light curve of eclipsing binary systems is known as the O'Connell effect (\citealt{1951PRCO....2...85O}). The presence of starspot(s) caused by the magnetic activity of the components is the most acceptable explanation for this phenomenon (\citealt{2017AJ....153..231S}). The light curves of the systems in all observed filters indicate the presence of the O'Connell effect, whereby asymmetry in maxima is clearly visible. So, the light curve solutions required spots accounting for the O'Connell effect and the observed light curve asymmetries. Figure \ref{Fig5} depicts the geometrical structure of these systems along with the spots that are displayed on the stars.

\vspace{1.5cm}
\section{Absolute parameters}\label{sec:absolute} 
We utilized Gaia EDR3's parallax and computed the systems distances to get a more satisfactory result for the absolute parameters. The Gaia EDR3 parallax was used to calculate the distances of the systems, and $V_{system}$ were obtained using the observed light curves to estimate the value of the $M_{v(system)}$. We used the extinction coefficients $(A_v)$ given by \cite{2011ApJ...737..103S}, assuming $R_v=3.1$. We estimated $M_{v1}$ and $M_{v2}$ for each system by considering $l_1/l_{tot}$ and $l_2/l_{tot}$ from our light curve solutions. The bolometric absolute magnitude for each component was calculated using the well-known equation $M_{bol}=M_v+BC$. The bolometric correction ($BC$) was calculated using \cite{1996ApJ...469..355F}'s table which was then improved by \cite{2010AJ....140.1158T}. Furthermore, $L_1$, $L_2$, $R_1$, and $R_2$ can be calculated for each component, respectively. We estimated $a_1$ and $a_2$, as well as the average, using the well-known $R=a\times r$ relation. As a result, we used Kepler's third law and the mass ratio to estimate the masses of the components ($M$). Moreover, the relationship between the mass and radius of the stars was used to derive $log(g)$. The absolute parameter uncertainties were estimated using the error bars of the involved parameters. The \cite{2022MNRAS.510.5315P} study has further information and formulas for calculating the absolute parameters using Gaia EDR3 parallax for contact systems. Table \ref{tab5} contains the parameters used during the calculations and the estimates of the absolute parameters of four of the systems.

\begin{table*}
\caption{Estimated values for absolute parameters, and other elements used during calculations.}
\centering
\begin{center}
\footnotesize
\begin{tabular}{c c c c c}
 \hline
 \hline
Parameter & EI Oct	& V336 TrA	& NX Boo & V356 Boo \\
\hline
$M_{v1}(mag)$ & $5.667(48)$ & $6.674(60)$ & $7.350(27)$ & $5.735(60)$ \\
$M_{v2}(mag)$ & $5.303(48)$ & $7.317(60)$ & $5.832(25)$ & $5.722(59)$ \\
$M_{bol1}(mag)$ & $5.368(48)$ & $6.370(60)$ & $7.082(27)$ & $5.570(60)$ \\
$M_{bol2}(mag)$ & $5.004(48)$ & $7.013(60)$ & $5.564(25)$ & $5.557(59)$ \\
$L_1(L_{\odot})$ & $0.561(25)$ & $0.223(12)$ & $0.116(3)$ & $0.466(25)$ \\
$L_2(L_{\odot})$ & $0.784(34)$ & $0.123(6)$ & $0.468(11)$ & $0.471(25)$ \\
$R_1(R_{\odot})$ & $0.880_{\rm-(35)}^{+(95)}$ & $0.638_{\rm-(90)}^{+(75)}$ & $0.516_{\rm-(22)}^{+(22)}$ & $0.204_{\rm-(39)}^{+(39)}$ \\
$R_2(R_{\odot})$ & $1.093_{\rm-(44)}^{+(45)}$ & $0.488_{\rm-(74)}^{+(58)}$ & $0.910_{\rm-(172)}^{+(86)}$ & $0.255_{\rm-(9)}^{+(19)}$ \\
$M_1(M_{\odot})$ & $0.740_{\rm-(66)}^{+(36)}$ & $0.367_{\rm-(76)}^{+(30)}$ & $0.272_{\rm-(151)}^{+(105)}$ & $0.012_{\rm-(2)}^{+(2)}$ \\
$M_2(M_{\odot})$ & $1.237_{\rm-(146)}^{+(90)}$ & $0.210_{\rm-(57)}^{+(55)}$ & $0.843_{\rm-(656)}^{+(277)}$ & $0.016_{\rm-(4)}^{+(6)}$ \\
${log(g)_1}$ & $4.418_{\rm-(6)}^{+(69)}$ & $4.393_{\rm-(31)}^{+(63)}$ & $4.447_{\rm-(497)}^{+(106)}$ & $3.910_{\rm-(108)}^{+(71)}$ \\
${log(g)_2}$ & $4.453_{\rm-(19)}^{+(254)}$ & $4.383_{\rm-(6)}^{+(1)}$ & $4.446_{\rm-(472)}^{+(654)}$ & $3.840_{\rm-(94)}^{+(512)}$ \\
$a(R_{\odot})$ & $2.565_{\rm-(95)}^{+(53)}$ & $1.451_{\rm-(121)}^{+(68)}$ & $1.737_{\rm-(175)}^{+(85)}$ & $0.560_{\rm-(42)}^{+(50)}$ \\
\hline
$Gaia$ $d(pc)$ & $253.137(7.655)$ & $96.041(0.119)$ & $273.902(1.219)$ & $237.474(0.926)$ \\
$A_v$ & $0.243(1)$ & $0.070(2)$ & $0.032(1)$ & $0.018(2)$ \\
$Obs.V_{system}$ & $11.977(20)$ & $11.178(61)$ & $12.812(34)$ & $11.872(67)$ \\
$BC$ & $-0.299$ & $-0.304$ & $-0.268$ & $-0.165$ \\
\hline
\hline
\end{tabular}
\end{center}
\label{tab5}
\end{table*}

The two systems PS Boo and V2282 Cyg do not have a complete observational light curve and the primary and secondary minimums have been observed. A new ephemeris for each of the systems is presented in this study. The period-mass relations can be used to obtain the absolute parameters and the mass ratio of these systems. The \cite{2022MNRAS.510.5315P} study analyzed a sample of 118 W UMa systems in which the absolute parameters of the stars were recalculated based on the light curve solutions in the previous studies and that the Gaia parallax was used to obtain new orbital period-mass relationships.
Therefore, according to the orbital periods of these systems, we first calculated the mass of each component (Equations 2 and 3). We then calculated the value of $log(g)$ using the relationship between the orbital period and the surface gravity of the stars (Equations 4 and 5).

\begin{equation}\label{eq2}
M_1=(2.924\pm0.075)P+(0.147\pm0.029)
\end{equation}

\begin{equation}\label{eq3}
M_2=(0.541\pm0.092)P+(0.294\pm0.034)
\end{equation}

\begin{equation}\label{eq4}
log(g)_1=(-1.436\pm0.068)P+(4.914\pm0.025)
\end{equation}

\begin{equation}\label{eq5}
log(g)_2=(-1.329\pm0.044)P+(4.763\pm0.016)
\end{equation}

We also calculated the value of each of the stellar radii using the mass and $log(g)$ values with the well-known relation (Equation 6), where radius and mass are solar units.

\begin{equation}\label{eq6}
R=\sqrt{(G_\odot M/g)}
\end{equation}

The mass ratio value has been determined by using the calculations completed for the component mass values $(q=M_2/M_1)$. Also, from Kepler's third law, the value of $a(R_\odot)$ can be calculated by considering the orbital period in seconds where $G$ is the gravitational constant (Equation 7).

\begin{equation}\label{eq7}
\frac{a^3}{G(M_1+M_2)}=\frac{P^2}{4\pi^2}
\end{equation}

Estimated values for these systems are given in Table \ref{tab6}.

\begin{table*}
\caption{Estimated absolute parameters of two systems based on their orbital period.}
\centering
\begin{center}
\footnotesize
\begin{tabular}{c c c}
 \hline
 \hline
Parameter & PS Boo	& V2282 Cyg \\
\hline
$M_1(M_{\odot})$ & $0.971(50)$ & $1.129(54)$ \\
$M_2(M_{\odot})$ & $0.446(60)$ & $0.476(65)$ \\
$R_1(R_{\odot})$ & $0.823(41)$ & $1.147(71)$ \\
$R_2(R_{\odot})$ & $0.499(35)$ & $0.629(41)$ \\
${log(g)_1}$ & $4.509(44)$ & $4.431(48)$ \\
${log(g)_2}$ & $4.389(28)$ & $4.317(31)$ \\
$a(R_{\odot})$ & $2.030(50)$ & $2.381(59)$ \\
\hline
$q$ & $0.460(38)$ & $0.422(37)$ \\
\hline
\hline
\end{tabular}
\end{center}
\label{tab6}
\end{table*}

\vspace{1.5cm}
\section{Discussion of some results}\label{sec:absolute} 
The main results from our analysis for six systems are as follows:

(1) We used PHOEBE's $q$-search to find a mass ratio initial value for the following modeling processes. The final mass ratio values for each of the systems were then obtained using the MCMC approach. For systems EI Oct, NX Boo, and V356 Boo, the value of mass ratio was more than one, and for V336 TrA this value was around 0.57.

\cite{2020MNRAS.497.3493Z} presented formulations describing how the orbital period is related to the mass ratio using a sample of 370 contact binaries. The mass ratio in the \cite{2020MNRAS.497.3493Z} study is the ratio of the less massive component mass to the more massive component mass, hence $q$ is never more than unity, so, they demonstrate this with $q^*$. Table \ref{tab7} presents comparative results according to which the mass ratio values in this study are consistent with this model.

\begin{table*}
\caption{Estimated mass ratio with two methods. The lower and upper limits of $q^*$ are shown from the \cite{2020MNRAS.497.3493Z} study with $q_{\rm l}^{*}$ and $q_{\rm u}^{*}$, respectively.}
\centering
\begin{center}
\footnotesize
\begin{tabular}{c c c c c c c}
 \hline
 \hline
 & & This study & & & \cite{2020MNRAS.497.3493Z} \\
System & $q$ & $q^*$ & & $q^*$ & $q_{\rm l}^{*}$ & $q_{\rm u}^{*}$ \\
\hline
EI Oct & $1.671$ & 0.598 && $0.38$ & $0.02$ & $0.92$ \\
V336 TrA & $0.573$ & 0.573 && $0.47$ & $0.09$ & $0.97$ \\
NX Boo & $3.100$ & 0.323 && $0.51$ & $0.10$ & $0.99$ \\
V356 Boo & $1.330$ & 0.752 && $0.44$ & $0.05$ & $0.95$ \\
PS Boo & $0.460$ & 0.460 && $0.45$ & $0.06$ & $0.95$ \\
V2282 Cyg & $0.422$ & 0.422 && $0.38$ & $0.02$ & $0.92$ \\
\hline
\hline
\end{tabular}
\end{center}
\label{tab7}
\end{table*}

(2) The minimum and maximum temperatures of the EI Oct, V336 TrA, and NX Boo's components in the light curve solutions are between 4692 and 5330 K, which shows that the studied stars are all from the G-K spectral types. The maximum temperature difference between the components is less than 318 K. The temperature difference between two components in W UMa systems is roughly 5\%, and this is true for all of our four systems (\citealt{2021RAA....21..203P}).

We can compute the ($B-V$) index and get a decent estimate of a star's surface temperature by using observational data from the blue and visual filters (\citealt{2021RAA....21..203P}). We determined ($B-V$) to be $0^m.810\pm0.008$ and $0^m.863\pm 0.004$, respectively, for systems EI Oct and V336 TrA, based on our observations and after calibration (\citealt{2000A&A...357..367H}). As a result, the effective temperatures of the primary components were found to be $5248\pm52$ and $5050\pm61$ K for EI Oct and V336 TrA, respectively, according to the \cite{2018MNRAS.479.5491E} study. These temperatures agree with the values in Gaia DR2, indicating that the accuracy of the observations was acceptable. By combining the results of ($B-V$) with the study of \cite{2000AJ....120.1072S}, it can be concluded that the metallicity (Fe/H) ratings for both primary stars to be -0.75 and -0.25, respectively (stars are in population II).

\begin{figure}
\begin{center}
\includegraphics[width=\columnwidth]{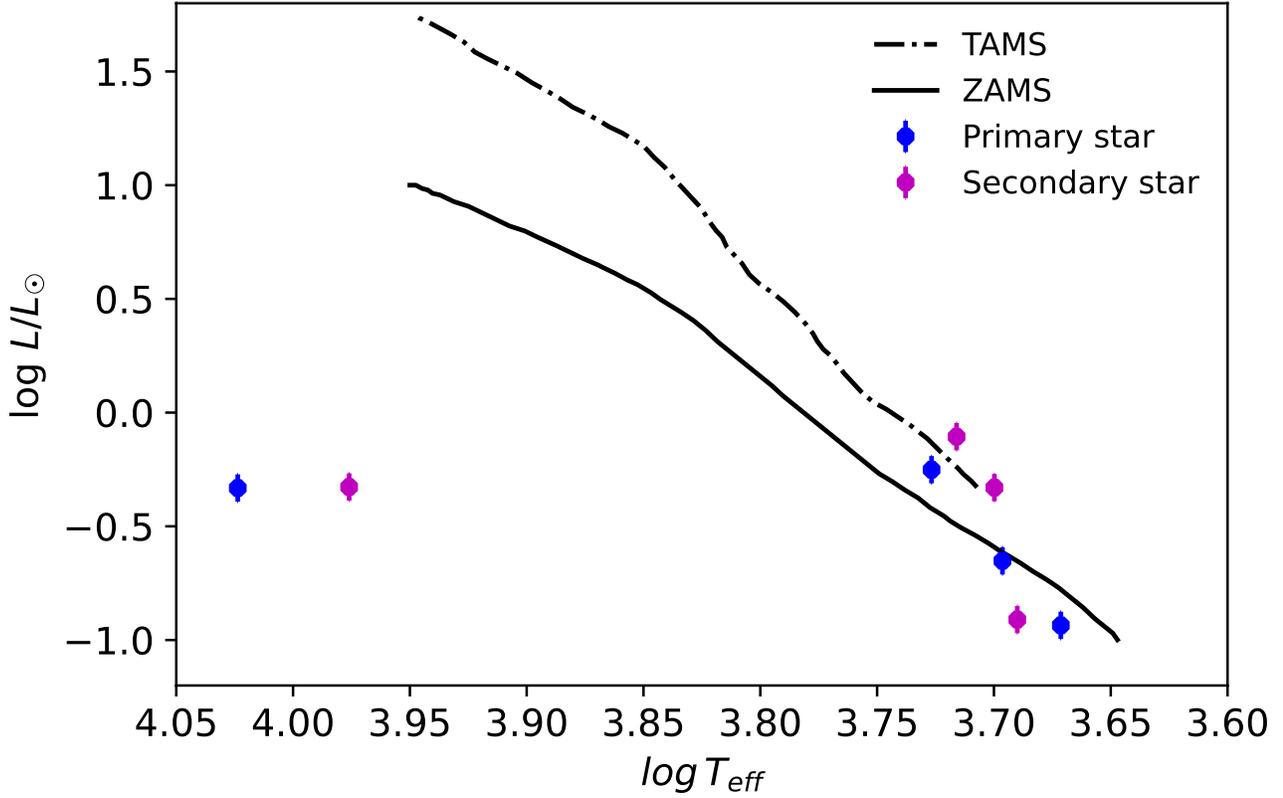}
\caption{The positions of the primary and secondary components in the H-R diagram in which the theoretical ZAMS and TAMS curves are indicated for EI Oct, V336 TrA, NX Boo, and V356 Boo systems.}
\label{Fig6}
\end{center}
\end{figure}

(3) We plotted the components of the systems in the H-R diagram for EI Oct, V336 TrA, NX Boo, and V356 Boo (Figure \ref{Fig6}).

The EI Oct system's primary star is in the main sequence, while the secondary star is above the main sequence's terminal age (TAMS). Moreover, the primary component of V336 TrA is in the main sequence, and the secondary star is below the theoretical zero-age main sequence (ZAMS). NX Boo's hotter star is in the main sequence, whereas the system's cooler star is below the ZAMS.

Furthermore, we propose that EI Oct is W-type W UMa binary systems based on mass ratio, the fillout factor, the inclination, and the absolute parameters. According to this, V336 TrA, and NX Boo are an A-type W UMa system.

The stars of the V356 Boo system are virtually in the region of the white dwarfs, as shown in the H-R diagram. It would only be normal for the components in this system to be outside the main sequence, given their temperature range of 9500–10500 K. The temperature specified in the Gaia DR2 for V356 Boo is 5402 K, which is very different from the temperature obtained in this study based on the MCMC method. The temperature difference between Gaia DR2 and MCMC, as well as the mass estimates of the stars in this system, suggest that the V356 Boo system will require spectroscopic investigation in the future.

(4) Asymmetry can be seen in the maxima of the light curves of EI Oct, V336 TrA, NX Boo, and V356 Boo systems. Therefore, the light curve solutions required cold or hot spots to account for the O'Connell effect.

(5) V336 TrA was analyzed by \cite{2018NewA...61....1K} for the first time. They obtained $q=1.396(3)$, $f=15.69(78)$, $i=80.80(11)$, $T_1=5000$, $T_2=4840(13)$ in their light curve solutions, and $M_1=0.653$, $M_2=0.912$, $R_1=0.732$, $R_2=0.850$, $L_1=0.300$, and $L_2=0.355$ in the calculation of the absolute parameters. Our results are completely different from the \cite{2018NewA...61....1K} study. The most important difference is the mass ratio value that we have obtained as 0.573.
\cite{2018NewA...61....1K} calculated the mass of each component by the mass ratio and the orbital period relationship. \cite{2018NewA...61....1K} concluded that V336 TrA is a W-type contact binary system which we cannot confirm it.

\begin{figure}
\begin{center}
\includegraphics[width=\columnwidth]{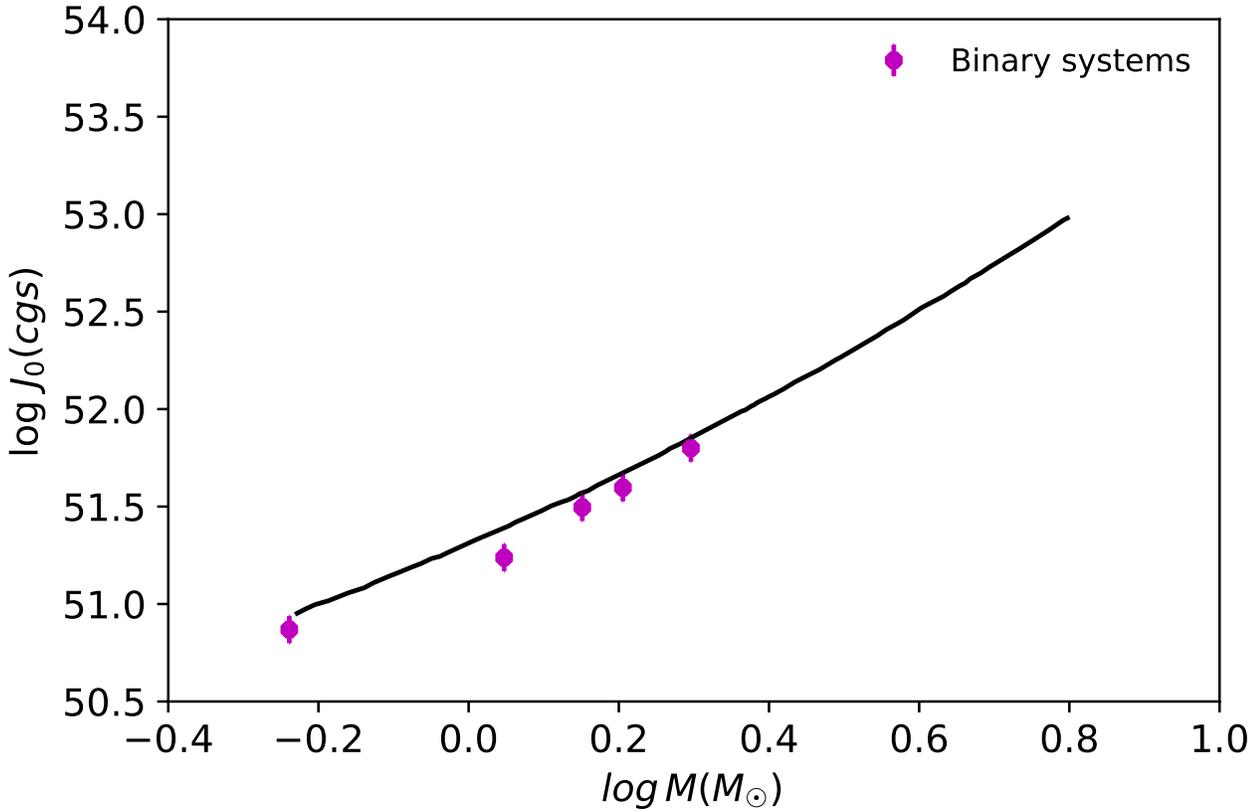}
\caption{The positions of the EI Oct, V336 TrA, NX Boo, PS Boo, and V2282 Cyg systems on the $logJ_0-logM$ scale. Uncertainty values are displayed as averages for these systems. The quadratic line is based on a study by \cite{2006MNRAS.373.1483E}.}
\label{Fig7}
\end{center}
\end{figure}

(6) We calculated the orbital angular momentum ($J_0$) of the six systems. Therefore, the value of $logJ_0$ for EI Oct=$51.800$, V336 TrA=$50.869$, NX Boo=$51.238$, V356 Boo=$48.713$, PS Boo=$51.496$ and V2282 Cyg=$51.597$, respectively. These results are based on the equation provided by \cite{2006MNRAS.373.1483E} as follows:

\begin{equation}\label{eq8}
J_0=\frac{q}{(1+q)^2}(\sqrt[3] {\frac{G^2}{2\pi})}M^5P
\end{equation}

where $q$ is the mass ratio, $M$ is the total mass of the components, and $P$ is the orbital period. We displayed the results in the $logJ_0-logM$ diagram and determined the location of each system (Figure \ref{Fig7}). The area above this quadratic line corresponds to detached systems and the area below it corresponds to contact binary systems. As a result, all studied systems are below the quadratic-fit and in the contact binary region.

\vspace{1.5cm}
\section{$M-R$ relationships for contact systems}\label{sec:MR} 
\cite{2011ApJ...728...48K} defined the principal stellar parameters with a critical influence on the evolution and internal structure of stars as well as in testing the stellar models. The Mass-Radius Relation (MRR) has been investigated significantly because of its numerous applications in astrophysics and the possibility to compare it with theoretical predictions. The practical purpose of MRR is to estimate the typical radius of main-sequence (MS) stars of a given mass. The MRR has been investigated in various types of stars; However, the focus on contact binary systems have been enhanced in recent decades.

In the study of non-contact binary systems, \cite{1977ApJS...34..479L} and \cite{1985Ap&SS.114..259G} derived two linear MRRs for two ranges of masses. \cite{2007MNRAS.382.1073M} used polynomials to calculate MRR with a mass and radius uncertainty of less than 10\%; Then, \cite{2018MNRAS.479.5491E} interrelated the relations of mass with luminosity, radius and effective temperature by studying a sample of 509 detached eclipsing binaries.

Some studies focused on samples containing various types of eclipsing binaries including contact systems. \cite{1981A&AS...46..193H} obtained empirical mass-radius relations for the MS components of eclipsing binaries. Then, \cite{1991Ap&SS.181..313D} obtained well-defined MRRs with a change in the slope around $M\approx1.7M_\odot$. The study performed by \cite{1998ARep...42..793G} resulted in the MRR of 112 eclipsing binary systems with both components on the MS by engaging the least-square method.

The contact binary systems were studied independently in numerous investigations of the MRR. \cite{2000SouSt..38..201K}, in an orbital period behavior study of 11 W-type and eight A-type systems calculated an MRR which was different from the relation of those stars located on the ZAMS line. Then, \cite{2005JKAS...38...43A}, by adopting a sample of 80 contact binaries (42 W-type and 38 A-type), determined mass-radius relations for both the subtypes separately. In the investigation of the mass evolution by \cite{2008MNRAS.390.1577G}, a power-law fit was derived from 112 contact binaries that were similar to the relation derived by \cite{1985Ap&SS.114..259G} for single MS stars with $M<1.8 M_\odot$. Their result declared that unlike secondary components, the primary components of cool contact binaries follow the MRR of MS stars. Another study by \cite{2018PASA...35....8K} found that the calculated mass-luminosity relation is not as precise as the mass-radius relation in the selected sample of 10 ultrashort-period overcontact binaries. \cite{2020PASJ...72..103L} in a statistical study of 380 Kepler contact binary systems found that the linear MRR of the primary and secondary stars is identical and that the primary stars are found generally on the ZAMS line. Then, \cite{2020MNRAS.492.4112Z} analyzed secondary components of 48 A-type and 69 W-type contact binaries in a mass-relation diagram and derived a relation based on radial density distribution that differs from the relation of MS stars. Also \cite{2020ApJ...905...39S}, in the investigation of energy transfer and its influence on the evolution of contact binaries, studied 70 W-type W UMa contact binaries derived from \cite{2009MNRAS.396.2176J}. They found that despite the primary components following the empirical mass-radius relation of the single stars, the secondary components are oversized and do not follow a similar relation. Eventually, \cite{2021ApJS..254...10L} derived linear mass-radius relations for the primary and secondary of 277 systems as well as for MS single stars. The diagram demonstrates that the primary components and unevolved low-mass single stars are located in a similar area. As a result of energy exchange through the common envelope the size of the secondary components becomes considerably larger than their masses (\citealt{2021ApJS..254...10L}; \citealt{2020ApJ...905...39S}).
Table \ref{tab8} shows the results of the previous studies comprising measured $M-R$ relationships for the contact binary systems.

\begin{table*}
\caption{Relations between mass and radius of components for EW-type systems based on the previous studies.}
\centering
\begin{center}
\footnotesize
\begin{tabular}{c c c c }
 \hline
 \hline
Parameters & Relation & Comment & Reference \\
\hline
$M-R$ & $R\propto M^{0.41}$ & & \cite{2000SouSt..38..201K} \\
$M-R$ & $R=0.33+0.55M$ & & \cite{2018PASA...35....8K} \\
\hline
$M_1-R_1$ & $log(R_{1W})=0.62log(M_{1W})+0.02$ & & \cite{2005JKAS...38...43A} \\
$M_1-R_1$ & $log(R_{1A})=0.62log(M_{1A})+0.09$ & & \cite{2005JKAS...38...43A} \\
$M_1-R_1$ & ${R_1}\propto M_1^{(0.92\pm0.04)}$ & & \cite{2008MNRAS.390.1577G} \\
$M_1-R_1$ & ${R_1}\propto M_1^{(0.90\pm0.03)}$ & & \cite{2020ApJ...905...39S} \\
$M_1-R_1$ & $logR_1=(0.90\pm0.03)logM_1+(0.04\pm0.01)$ & & \cite{2021ApJS..254...10L} \\
\hline
$M_2-R_2$ & $log(R_{2W})=0.44log(M_{2W})+0.02$ & & \cite{2005JKAS...38...43A} \\
$M_2-R_2$ & $log(R_{2A})=0.31log(M_{2A})+0.07$ & & \cite{2005JKAS...38...43A} \\
$M_2-R_2$ & $M_2=4\pi/(3+\beta)\delta \times R_{\rm2}^{3+\beta}$ & ${\delta=0.077}, {\beta=-1.227}, {M_2\geqslant1.8M_\odot}$ & \cite{2020MNRAS.492.4112Z} \\
$M_2-R_2$ & $M_2=4\pi/(3+\beta)\delta \times R_{\rm2}^{3+\beta}$ & $\delta=0.071, \beta=-2.004, {M_2\leqslant1.8M_\odot}$ & \cite{2020MNRAS.492.4112Z} \\
$M_2-R_2$ &  $logR_2=(0.38\pm0.03)logM_2+(0.06\pm0.01)$ & & \cite{2021ApJS..254...10L} \\
\hline
\hline
\end{tabular}
\end{center}
\label{tab8}
\end{table*}

We used a sample of 118 contact binary systems from the \cite{2022MNRAS.510.5315P} study to investigate the $M-R$ relationships in contact systems. In this sample, the Gaia EDR3 parallax was used to calculate the absolute parameters for the selected systems with orbital periods shorter than 0.6 days. \cite{2022MNRAS.510.5315P} study focused on the relationships between orbital period and mass, and we used a similar method to calculate the absolute parameters in Section 5 for the six systems studied.

\begin{figure}
\begin{center}
\includegraphics[width=\columnwidth]{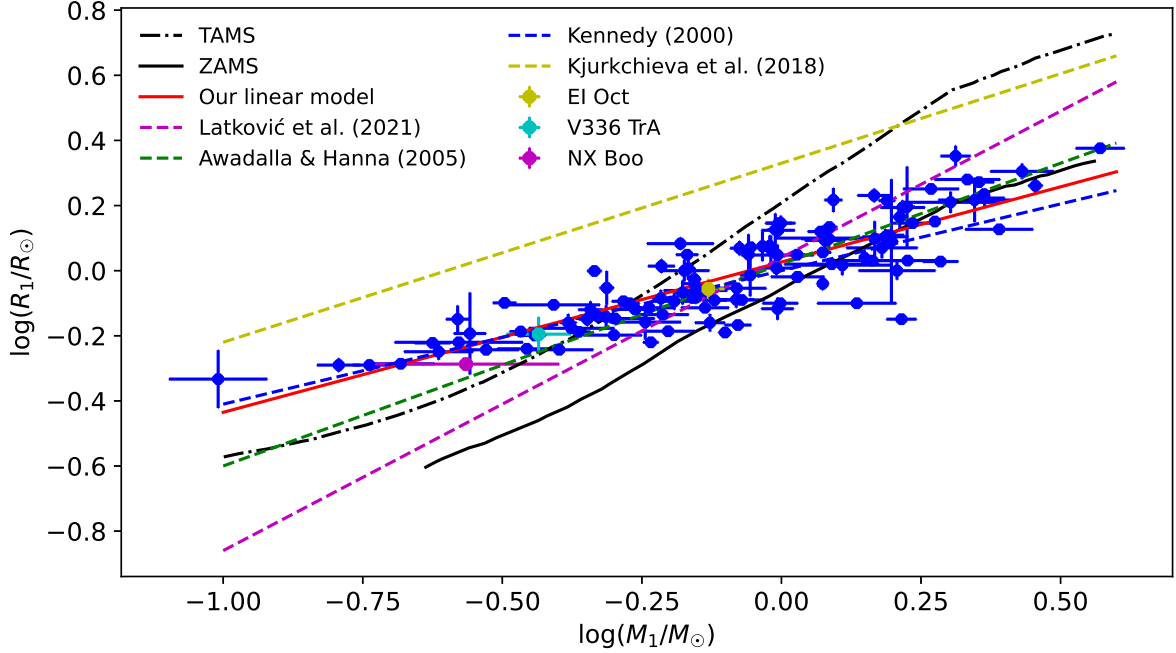}
\caption{The $M_1-R_1$ relationship diagram for contact binary systems. The position of the EI Oct, V336 TrA, and NX Boo systems that were analyzed in this study is also shown in the figure in different colors.}
\label{Fig8}
\end{center}
\end{figure}

\begin{figure}
\begin{center}
\includegraphics[width=\columnwidth]{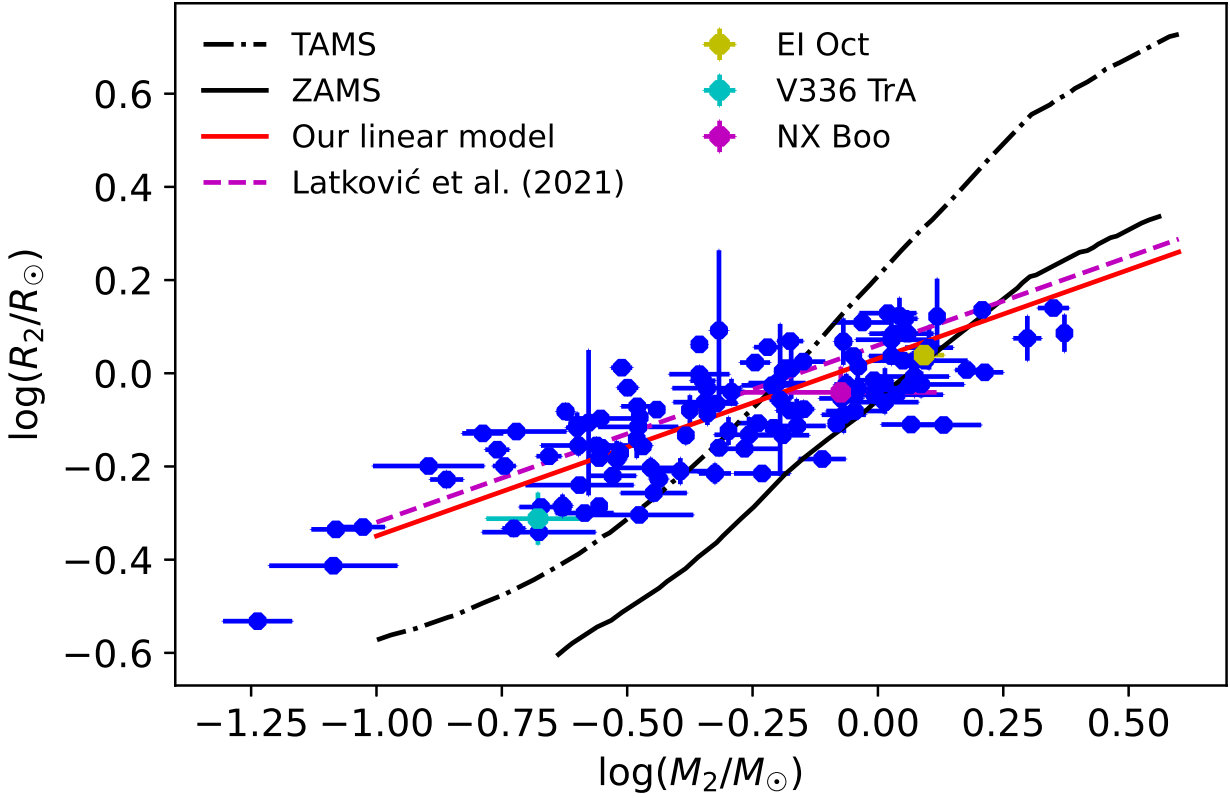}
\caption{The $M_2-R_2$ relationship diagram for contact binary systems. The results of the photometric analysis of the EI Oct, V336 TrA, and NX Boo systems in this investigation is shown.}
\label{Fig9}
\end{center}
\end{figure}

\begin{figure}
\begin{center}
\includegraphics[width=\columnwidth]{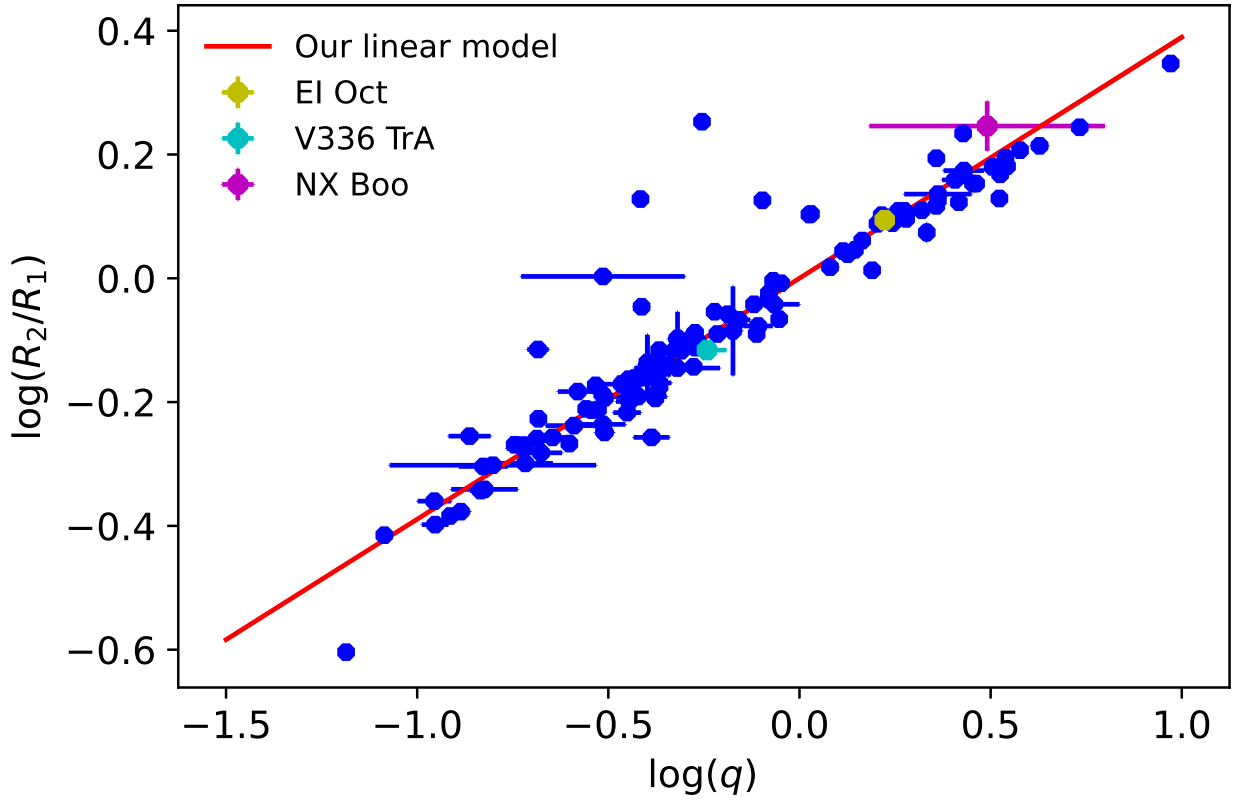}
\caption{The mass ratio and radius ratio relationship diagram for contact binary systems. The estimated mass ratios from the light curve solutions of the EI Oct, V336 TrA, and NX Boo systems in this study, and the results of the mass ratio calculated absolute parameters are displayed for comparison with the theoretical linear model.}
\label{Fig10}
\end{center}
\end{figure}

\begin{figure*}
\begin{center}
\includegraphics[width=15cm,height=5cm]{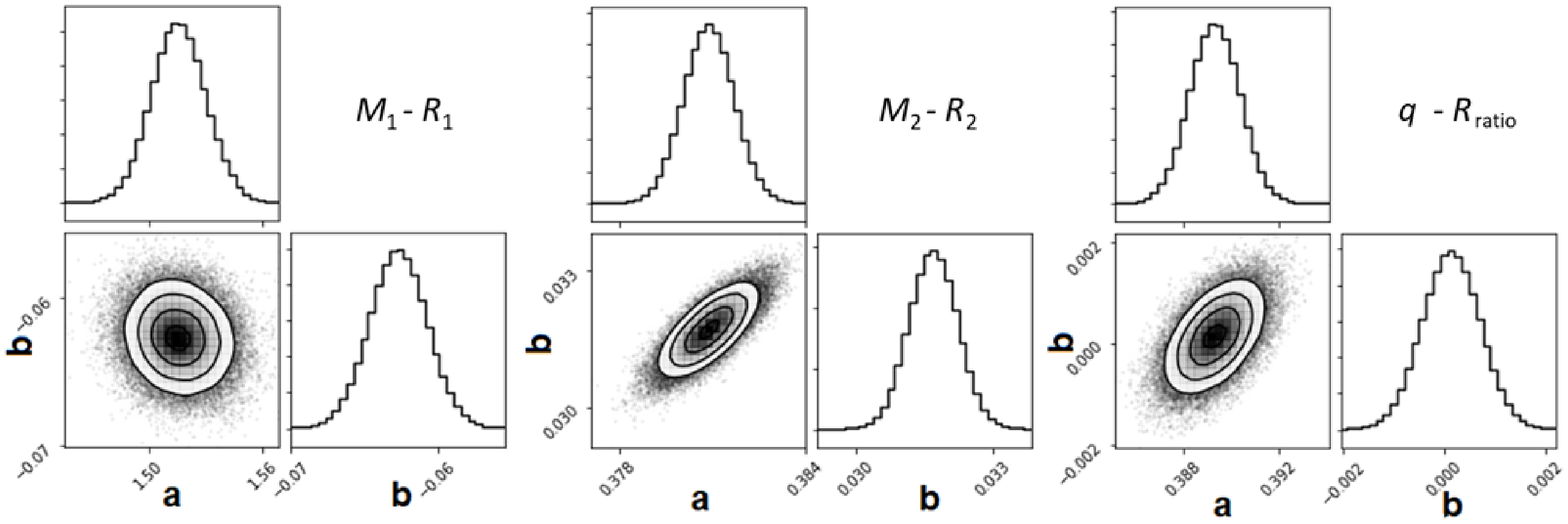}
\caption{Corner plots of the posterior distribution based on the MCMC sampling (a and b are the fitted parameters of the linear models on the data).}
\label{Fig11}
\end{center}
\end{figure*}

We applied the models between physical quantities of $R$, $M$, and their relationships (Figures \ref{Fig8}, \ref{Fig9}, and \ref{Fig10}). There are three linear models between mass and radius ($M_1-R_1 and M_2-R_2$), and a linear model between mass ratio ($q=M_2/M_1$) and radius ratio ($R_2/R_1$) of contact binary systems. By the MCMC sampling, we found the final distribution of the model's parameters and their uncertainties. For this objective, we employed the PyMC3 package (\citealt{2016ascl.soft10016S}) in Python, which this sampling ran for 20 walkers, 10000 iterations, and 1000 burn-in. Figure \ref{Fig11} displays the corner plots of the posterior distribution for $M_1-R_1$, $M_2-R_2$, and mass ratio-radius ratio based on MCMC sampling.
The two new relationships between the mass and radius of the contact binary’s components are presented in the following equations 9 and 10,

\begin{equation}\label{eq9}
logR_1=(0.4617\pm0.0250)logM_1+(0.0266\pm0.0076)
\end{equation}

\begin{equation}\label{eq10}
logR_2=(0.3808\pm0.0008)logM_2+(0.0317\pm0.0005)
\end{equation}

In addition, Equation 11 shows a new relationship between the mass ratio and the radius ratio in the contact binary systems.

\begin{equation}\label{eq11}
log(\frac{R_2}{R_1} )=(0.3893\pm0.0010)logq+(0.0001\pm0.0005)
\end{equation}

Figure \ref{Fig8} shows some of the relationships presented in the previous studies and in this study. As can be seen in Figure \ref{Fig8} and Table \ref{tab8}, there are relatively large differences between some of the relationships for $M_1-R_1$. In the case of Figure \ref{Fig9} for the $M_2-R_2$ relationships, this difference is small. This discrepancy seems to be more due to the variety of the selected samples. Given that our selected sample (118 contact systems) was calculated by the Gaia EDR3 parallax method (\citealt{2022MNRAS.510.5315P}) and, unlike most previous studies, was not collected from other studies with different methods, it seems significant. To further compare the relationships presented in this study and the \cite{2021ApJS..254...10L} study, we measured them using the results of four studies (Table \ref{tab9}). In these studies, which investigated seven systems, the results were based on spectroscopy.
The \cite{2021AJ....161..221P} study, which investigated four W UMa systems (J015829.5+260333 (J0158), J030505.1+293443 (J0305), J102211.7+310022 (J1022), and KW Psc) used spectroscopic results, light curve analysis, and Gaia parallax to obtain the radius values.

In the studies by \cite{2011AN....332..690G}, \cite{2016NewA...47...57G}, and \cite{2015NewA...36..100G}, they used radial velocity, light curve solutions, and the \cite{1983ApJ...268..368E} relation for calculating the radii of each component.

\begin{table*}
\caption{Comparison of this study (model 1) and the \cite{2021ApJS..254...10L} study (model 2) on $M-R$ relations, with the result of seven systems based on spectroscopy.}
\centering
\begin{center}
\footnotesize
\begin{tabular}{c c c c c c c c}
\hline
\hline
Parameters & V404 Peg & V1918 Cyg & V546 And & J0158 & J0305 & J1022 & KW Psc \\
\hline
$M_1$ & $1.175(25)$ & $1.302(69)$ & $0.275(8)$ & $1.262(171)$ & $0.927(85)$ & $0.313(27)$ & $0.557(53)$ \\
$M_2$ & $0.286(6)$ & $0.362(19)$ & $1.083(30)$ & $0.846(189)$ & $0.287(28)$ & $1.011(97)$ & $0.234(23)$ \\
$R_1$ & $1.346(10)$ & $1.362(24)$ & $0.661(6)$ & $1.381(53)$ & $0.865(26)$ & $0.547(14)$ & $0.694(22)$ \\
$R_2$ & $0.710(5)$ & $0.762(19)$ & $1.229(16)$ & $1.259(48)$ & $0.509(16)$ & $0.929(24)$ & $0.474(15)$ \\
Reference & 1\footnote{\cite{2011AN....332..690G}} &  2\footnote{\cite{2016NewA...47...57G}} &  3\footnote{\cite{2015NewA...36..100G}} & 4\footnote{\cite{2021AJ....161..221P}} & 4\footnote{\cite{2021AJ....161..221P}} & 4\footnote{\cite{2021AJ....161..221P}} & 4\footnote{\cite{2021AJ....161..221P}} \\
\hline
$R_1$(Model1) & $1.145(11)$ & $1.201(29)$ & $0.586(8)$ & $1.184(72)$ & $1.027(42)$ & $0.622(24)$ & $0.811(35)$ \\
$R_1$(Model2) & $1.268(24)$ & $1.390(66)$ & $0.343(9)$ & $1.352(64)$ & $1.024(84)$ & $0.385(30)$ & $0.648(55)$ \\
$R_2$(Model1) & $0.668(5)$ & $0.731(14)$ & $1.109(12)$ & $1.009(81)$ & $0.669(24)$ & $1.080(38)$ & $0.619(22)$ \\
$R_2$(Model2) & $0.714(6)$ & $0.780(15)$ & $1.183(12)$ & $1.077(86)$ & $0.714(26)$ & $1.153(41)$ & $0.661(24)$ \\
\hline
\hline
\end{tabular}
\end{center}
\label{tab9}
\end{table*}

\vspace{1.5cm}
\section{Summary and conclusion}\label{sec:conclusion} 
We performed the photometric analysis on six W UMa-type systems. These short-period systems were observed for 27 nights in various filters at three observatories in the northern and southern hemispheres. For these six systems, 37 minima times were observed and extracted and along with other minimum times from the literature, a new ephemeris was calculated using the MCMC approach for each system. The small number and short baseline of observations, as well as the type of trend found in all O-C diagrams, suggest that all six systems should be followed up in the future. Also, more observations are essential for the two PS Boo and V336 TrA systems, which may be candidates for the orbital growth. The orbital periods of the studied systems are 0.25 to 0.34 days.
EI Oct, V336 TrA, NX Boo, and V356 Boo systems had complete observational light curves. PHOEBE 2.3.59 version and the MCMC approach Python codes were used to analyze the light curves of these four systems. The systems' absolute parameters were estimated using their Gaia EDR3 parallax. For systems PS Boo and V2282 Cyg, each had observations of the primary and secondary minima, so some of their main absolute parameters and mass ratio were calculated by using the \cite{2022MNRAS.510.5315P} study equations.
Understanding the evolutionary state of the W UMa-type contact systems through the investigation of the relations between the fundamental stellar parameters such as mass and radius has become possible in the past decades. Multiple studies on the MRR have gradually improved our understanding of the evolutionary paths between A-type and W-type systems, as well as the stellar properties of the primary and secondary components.
For decades, the increasing studies on this subject have considered various ranges of the orbital period. Hence, the detached, semi-detached, and contact eclipsing binaries have been investigated individually or in combined samples. Also, in some papers, the two sub-types of contact binaries (A-type and W-type) have been studied separately.
In the present investigation, we used a sample of 118 contact binaries to calculate the absolute parameters using Gaia EDR3 parallaxes. By applying the linear model to our sample we updated the relations between $M-R_{(1,2)}$ and determined a new relation for radius ratio-mass ratio. The results of the  EI Oct, V336 TrA, and NX Boo systems analyzed in this study based on observational light curves are also shown in Figures \ref{Fig8}, \ref{Fig9}, and \ref{Fig10}, indicating that they are in good agreement with theoretical predictions.

We chose all 1473 binaries with a maximum confidence of EW-type (Classification probability=1) and an orbital period of less than 0.6 days from the ASAS-SN catalog. We estimated some of the absolute parameters (e.g. $M,R,q,a,log g$) of the components of this sample. The mass of the systems was determined using Equations 2 and 3 in this work (\citealt{2022MNRAS.510.5315P}); the radius of the stars and the quantity of the mass ratio\footnote{The value of $q^*$ is obtained by $M_2/M_1$, and the value of $q^{**}$ was calculated based on Equation 11 in this study.}  were computed using the new relationships of this study (Equations 9, 10, and 11); the value of the surface gravity was also computed and Kepler's third law was used to estimate the value of $a(R_\odot)$. The results are available as a machine-readable table in a supplement to this work\footnote{\url{https://iopscience.iop.org/journal/1538-3873}}.

\vspace{1.5cm}
\section*{Acknowledgments}
{This manuscript was prepared by the Binary Systems of South and North Project (\url{http://bsnp.info}). The goal of the project is to investigate contact binary systems in the northern and southern hemispheres from a variety of observatories throughout the world.
The SIMBAD database (\url{http://simbad.u-strasbg.fr/simbad/}), which is maintained by CDS in Strasbourg, France, was used in this study. We have made use of data from the European Space Agency (ESA) mission Gaia (\url{http://www.cosmos.esa.int/gaia}), processed by the Gaia Data Processing and Analysis Consortium (DPAC, \url{http://www.cosmos.esa.int/web/gaia/dpac/consortium}).
Thanks to the Raderon Lab AI department (\url{https://raderonlab.ca}) for providing the hardware needed for the MCMC process and support. We appreciate Dr. Fahri Alicavus' scientific cooperation. We also appreciate Elahe Lashgari's assistance. Thanks as well to Paul D. Maley and Tabassom Madayen for making editorial corrections to the text. The authors wish to thank Dr. Angela Kochoska for all of her help and advice in analyzing the light curves with the PHOEBE and MCMC.}

\clearpage
\appendix
Tables \ref{A1}, \ref{A2}, \ref{A3}, \ref{A4}, \ref{A5} and \ref{A6} show the minima times in the first column, the minimum time errors in the second column, the epochs in the third column, the O-C values in the fourth column, and the references in the last column.

\setcounter{table}{0}
\begin{table}[ht]
\caption{Available times of minima for EI Oct.}
\centering
\begin{center}
\footnotesize
\begin{tabular}{c c c c c }
\hline
\hline
Min.($BJD_{TDB}$)& Error & Epoch & O-C & Reference \\
\hline
$2454556.50654$ &  & $0$ & $0$ & \cite{2006SASS...25...47W} \\
$2459063.96445$	& $0.00010$ & $13315$ & $0.0042$ & This study \\
$2459064.13451$	& $0.00010$ & $13315.5$ & $0.0050$ & This study \\
$2459078.01393$	& $0.00047$ & $13356.5$	& $0.0049$ & This study \\
$2459078.01447$	& $0.00030$ & $13356.5$	& $0.0054$ & This study \\
$2459078.01456$	& $0.00017$ & $13356.5$	& $0.0055$ & This study \\
$2459078.18269$	& $0.00013$ & $13357$ & $0.0044$ & This study \\
$2459078.18315$	& $0.00028$ & $13357$ & $0.0049$ & This study \\
$2459078.18323$	& $0.00178$ & $13357$ & $0.0049$ & This study \\
$2459085.96978$	& $0.00014$ & $13380$ & $0.0054$ & This study \\
$2459086.13883$	& $0.00014$ & $13380.5$	& $0.0052$ & This study \\
$2459088.17075$	& $0.00013$ & $13386.5$	& $0.0060$ & This study \\
\hline
\hline
\end{tabular}
\end{center}
\label{A1}
\end{table}

\begin{table}[ht]
\caption{Available times of minima for V336 TrA.}
\centering
\begin{center}
\footnotesize
\begin{tabular}{c c c c c }
\hline
\hline
Min.($BJD_{TDB}$)& Error & Epoch & O-C & Reference \\
\hline
$	2452151.54277	$	&				&	$	0	$	&	$	0	$	&	\cite{2004IBVS.5600...27S}	\\
$	2456490.65369	$	&				&	$	16265.5	$	&	$	-0.004	$	&	\cite{2018NewA...61....1K}	\\
$	2456491.72019	$	&				&	$	16269.5	$	&	$	-0.0046	$	&	\cite{2018NewA...61....1K}	\\
$	2456492.65459	$	&				&	$	16273	$	&	$	-0.0038	$	&	\cite{2018NewA...61....1K}	\\
$	2456494.65479	$	&				&	$	16280.5	$	&	$	-0.0044	$	&	\cite{2018NewA...61....1K}	\\
$	2456496.52469	$	&				&	$	16287.5	$	&	$	-0.0019	$	&	\cite{2018NewA...61....1K}	\\
$	2456496.65469	$	&				&	$	16288	$	&	$	-0.0053	$	&	\cite{2018NewA...61....1K}	\\
$	2456497.58939	$	&				&	$	16291.5	$	&	$	-0.0043	$	&	\cite{2018NewA...61....1K}	\\
$	2459001.07358	$	&	$	0.00008	$	&	$	25676	$	&	$	-0.0044	$	&	This study	\\
$	2459001.07360	$	&	$	0.00001	$	&	$	25676	$	&	$	-0.0043	$	&	This study	\\
$	2459001.07362	$	&	$	0.00011	$	&	$	25676	$	&	$	-0.0043	$	&	This study	\\
$	2459008.94319	$	&	$	0.00009	$	&	$	25705.5	$	&	$	-0.0044	$	&	This study	\\
$	2459008.94325	$	&	$	0.00013	$	&	$	25705.5	$	&	$	-0.0043	$	&	This study	\\
$	2459008.94342	$	&	$	0.00008	$	&	$	25705.5	$	&	$	-0.0042	$	&	This study	\\
$	2459009.07653	$	&	$	0.00013	$	&	$	25706	$	&	$	-0.0044	$	&	This study	\\
$	2459009.07656	$	&	$	0.00008	$	&	$	25706	$	&	$	-0.0044	$	&	This study	\\
$	2459009.07657	$	&	$	0.00005	$	&	$	25706	$	&	$	-0.0044	$	&	This study	\\
$	2459016.01227	$	&	$	0.00013	$	&	$	25732	$	&	$	-0.0047	$	&	This study	\\
$	2459016.01251	$	&	$	0.00006	$	&	$	25732	$	&	$	-0.0044	$	&	This study	\\
$	2459016.01262	$	&	$	0.00019	$	&	$	25732	$	&	$	-0.0043	$	&	This study	\\
$	2459016.14593	$	&	$	0.00009	$	&	$	25732.5	$	&	$	-0.0044	$	&	This study	\\
$	2459016.14609	$	&	$	0.00010	$	&	$	25732.5	$	&	$	-0.0042	$	&	This study	\\
$	2459016.14625	$	&	$	0.00011	$	&	$	25732.5	$	&	$	-0.0041	$	&	This study	\\
$	2459326.26457	$	&	$	0.00040	$	&	$	26895	$	&	$	-0.0036	$	&	This study	\\
$	2459367.07974	$	&	$	0.00060	$	&	$	27048	$	&	$	-0.0039	$	&	This study	\\
\hline
\hline
\end{tabular}
\end{center}
\label{A2}
\end{table}

\begin{table}[ht]
\caption{Available times of minima for NX Boo.}
\centering
\begin{center}
\footnotesize
\begin{tabular}{c c c c c }
\hline
\hline
Min.($BJD_{TDB}$)& Error & Epoch & O-C & Reference \\
\hline
$	2451578.83777	$	&				&	$	-19443.5	$	&	$	-0.0131	$	&	\cite{2014OEJV..162....1P}	\\
$	2454640.79675	$	&	$	0.01000	$	&	$	-7251	$	&	$	-0.0054	$	&	\cite{2014OEJV..162....1P}	\\
$	2456016.89448	$	&	$	0.00030	$	&	$	-1771.5	$	&	$	0.0036	$	&	\cite{2012IBVS.6029....1D}	\\
$	2456086.70888	$	&	$	0.00030	$	&	$	-1493.5	$	&	$	0.0027	$	&	\cite{2012IBVS.6029....1D}	\\
$	2456461.77478	$	&				&	$	0	$	&	$	0	$	&	\cite{2006SASS...25...47W}	\\
$	2456745.56048	$	&	$	0.00050	$	&	$	1130	$	&	$	0.0043	$	&	\cite{2015OEJV..168....1H}	\\
$	2456760.37648	$	&	$	0.00030	$	&	$	1189	$	&	$	0.0034	$	&	\cite{2015OEJV..168....1H}	\\
$	2456760.50238	$	&	$	0.00010	$	&	$	1189.5	$	&	$	0.0037	$	&	\cite{2015OEJV..168....1H}	\\
$	2456764.39918	$	&	$	0.00040	$	&	$	1205	$	&	$	0.0079	$	&	\cite{2015OEJV..168....1H}	\\
$	2456798.42438	$	&	$	0.00010	$	&	$	1340.5	$	&	$	0.0045	$	&	\cite{2015OEJV..168....1H}	\\
$	2456815.37627	$	&	$	0.00020	$	&	$	1408	$	&	$	0.0048	$	&	\cite{2015OEJV..168....1H}	\\
$	2456815.50157	$	&	$	0.00020	$	&	$	1408.5	$	&	$	0.0045	$	&	\cite{2015OEJV..168....1H}	\\
$	2457066.88817	$	&	$	0.00090	$	&	$	2409.5	$	&	$	0.0060	$	&	\cite{2016IBVS.6157....1H}	\\
$	2457121.38497	$	&	$	0.00030	$	&	$	2626.5	$	&	$	0.0067	$	&	\cite{2016IBVS.6157....1H}	\\
$	2457128.41607	$	&	$	0.00120	$	&	$	2654.5	$	&	$	0.0061	$	&	\cite{2017IBVS.6196....1H}	\\
$	2457128.54067	$	&	$	0.00450	$	&	$	2655	$	&	$	0.0051	$	&	\cite{2017IBVS.6196....1H}	\\
$	2457149.38497	$	&	$	0.00020	$	&	$	2738	$	&	$	0.0053	$	&	\cite{2017OEJV..179....1J}	\\
$	2457149.38657	$	&	$	0.00020	$	&	$	2738	$	&	$	0.0069	$	&	\cite{2017OEJV..179....1J}	\\
$	2457461.54557	$	&	$	0.00050	$	&	$	3981	$	&	$	0.0063	$	&	\cite{2017OEJV..179....1J}	\\
$	2457464.43607	$	&	$	0.00190	$	&	$	3992.5	$	&	$	0.0088	$	&	\cite{2017IBVS.6196....1H}	\\
$	2457464.55887	$	&	$	0.00160	$	&	$	3993	$	&	$	0.0060	$	&	\cite{2017IBVS.6196....1H}	\\
$	2458228.38709	$	&	$	0.00170	$	&	$	7034.5	$	&	$	0.0102	$	&	\cite{2018IBVS.6244....1P}	\\
$	2458228.51089	$	&	$	0.00200	$	&	$	7035	$	&	$	0.0084	$	&	\cite{2018IBVS.6244....1P}	\\
$	2458945.50210	$	&	$	0.00270	$	&	$	9890	$	&	$	0.0121	$	&	\cite{2021BAVJ...52....1P}	\\
$	2459035.28801	$	&	$	0.00092	$	&	$	10247.5	$	&	$	0.0176	$	&	This study	\\
\hline
\hline
\end{tabular}
\end{center}
\label{A3}
\end{table}

\begin{table}[ht]
\caption{Available times of minima for V356 Boo.}
\centering
\begin{center}
\footnotesize
\begin{tabular}{c c c c c }
\hline
\hline
Min.($BJD_{TDB}$)& Error & Epoch & O-C & Reference \\
\hline
$	2453447.87472	$	&				&	$	-118.5	$	&	$	0.0051	$	&	\cite{2009IBVS.5894....1D}	\\
$	2453448.01772	$	&				&	$	-118	$	&	$	0.0049	$	&	\cite{2014OEJV..162....1P}	\\
$	2453481.81772	$	&	$	0.01000	$	&	$	0	$	&	$	0	$	&	\cite{2014OEJV..162....1P}	\\
$	2454958.77877	$	&	$	0.00200	$	&	$	5155.5	$	&	$	0.0031	$	&	\cite{2009IBVS.5894....1D}	\\
$	2454958.92077	$	&	$	0.00300	$	&	$	5156	$	&	$	0.0019	$	&	\cite{2009IBVS.5894....1D}	\\
$	2455639.89238	$	&	$	0.00040	$	&	$	7533	$	&	$	0.0058	$	&	\cite{2011IBVS.5960....1D}	\\
$	2455695.75839	$	&	$	0.00030	$	&	$	7728	$	&	$	0.0078	$	&	\cite{2011IBVS.5960....1D}	\\
$	2455695.89869	$	&	$	0.00040	$	&	$	7728.5	$	&	$	0.0048	$	&	\cite{2011IBVS.5960....1D}	\\
$	2456009.88598	$	&	$	0.00160	$	&	$	8824.5	$	&	$	0.0079	$	&	\cite{2012IBVS.6029....1D}	\\
$	2456089.67188	$	&	$	0.00010	$	&	$	9103	$	&	$	0.0085	$	&	\cite{2012IBVS.6029....1D}	\\
$	2457914.72068	$	&	$	0.00030	$	&	$	15473.5	$	&	$	0.0237	$	&	\cite{2018IBVS.6234....1N}	\\
$	2459026.41497	$	&	$	0.00052	$	&	$	19354	$	&	$	0.0246	$	&	This study	\\
$	2459027.41871	$	&	$	0.00086	$	&	$	19357.5	$	&	$	0.0257	$	&	This study	\\
\hline
\hline
\end{tabular}
\end{center}
\label{A4}
\end{table}

\begin{table}[ht]
\caption{Available times of minima for PS Boo.}
\centering
\begin{center}
\footnotesize
\begin{tabular}{c c c c c }
\hline
\hline
Min.($BJD_{TDB}$)& Error & Epoch & O-C & Reference \\
\hline
$	2451403.82576	$	&				&	$	0	$	&	$	0	$	&	\cite{2006PZP.....6...16K}	\\
$	2456725.55438	$	&	$	0.00040	$	&	$	18895.5	$	&	$	0.0038	$	&	\cite{2015OEJV..168....1H}	\\
$	2456745.41018	$	&	$	0.00190	$	&	$	18966	$	&	$	0.0040	$	&	\cite{2015IBVS.6149....1H}	\\
$	2456745.55228	$	&	$	0.00080	$	&	$	18966.5	$	&	$	0.0053	$	&	\cite{2015IBVS.6149....1H}	\\
$	2457116.47477	$	&	$	0.00130	$	&	$	20283.5	$	&	$	0.0081	$	&	\cite{2017OEJV..179....1J}	\\
$	2457465.56538	$	&	$	0.00040	$	&	$	21523	$	&	$	0.0062	$	&	\cite{2017OEJV..179....1J}	\\
$	2457878.46069	$	&	&	$22989$	&	$0.0176$	& B.R.N.O.\footnote{\url{http://var2.astro.cz/brno/}}	\\
$	2457927.45809	$	&	&	$23163$	&	$0.0096$	& B.R.N.O.	\\
$	2459005.31275	$	&	$	0.00190	$	&	$	26990	$	&	$	0.0288	$	&	This study	\\
$	2459005.44330	$	&	$	0.00164	$	&	$	26990.5	$	&	$	0.0185	$	&	This study	\\
\hline
\hline
\end{tabular}
\end{center}
\label{A5}
\end{table}

\begin{table}[ht]
\caption{Available times of minima for V2282 Cyg.}
\centering
\begin{center}
\footnotesize
\begin{tabular}{c c c c c }
\hline
\hline
Min.($BJD_{TDB}$)& Error & Epoch & O-C & Reference \\
\hline
$	2451258.83643	$	&	$	0.00060	$	&	$	-3694.5	$	&	$	-0.0050	$	&	\cite{2000IBVS.4996....1B}	\\
$	2451308.72784	$	&	$	0.00050	$	&	$	-3546	$	&	$	-0.0022	$	&	\cite{2000IBVS.4996....1B}	\\
$	2451801.40084	$	&				&	$	-2079.5	$	&	$	0.0001	$	&	\cite{2001BBSAG.124....D}	\\
$	2451805.44164	$	&				&	$	-2067.5	$	&	$	0.0095	$	&	\cite{2001BBSAG.124....D}	\\
$	2451809.47014	$	&				&	$	-2055.5	$	&	$	0.0066	$	&	\cite{2001BBSAG.124....D}	\\
$	2451811.31084	$	&				&	$	-2050	$	&	$	-0.0004	$	&	\cite{2001BBSAG.124....D}	\\
$	2451811.31314	$	&	$	0.00090	$	&	$	-2050	$	&	$	0.0019	$	&	\cite{2000IBVS.4996....1B}	\\
$	2451811.48394	$	&				&	$	-2049.5	$	&	$	0.0047	$	&	\cite{2001BBSAG.124....D}	\\
$	2451814.33464	$	&				&	$	-2041	$	&	$	-0.0002	$	&	\cite{2001BBSAG.124....D}	\\
$	2451850.28254	$	&				&	$	-1934	$	&	$	0.0011	$	&	\cite{2001BBSAG.124....D}	\\
$	2452112.48635	$	&				&	$	-1153.5	$	&	$	-0.0041	$	&	\cite{2000IBVS.4996....1B}	\\
$	2452443.40275	$	&				&	$	-168.5	$	&	$	0.0016	$	&	\cite{2000IBVS.4996....1B}	\\
$	2452500.00875	$	&				&	$	0	$	&	$	0	$	&	\cite{2006SASS...25...47W}	\\
$	2452820.50055	$	&	$	0.00070	$	&	$	954	$	&	$	-0.0045	$	&	\cite{2004IBVS.5543....1D}	\\
$	2453233.38065	$	&	$	0.00090	$	&	$	2183	$	&	$	-0.0069	$	&	\cite{2005IBVS.5653....1D}	\\
$	2453233.38065	$	&	$	0.00090	$	&	$	2183	$	&	$	-0.0069	$	&	\cite{2005IBVS.5653....1D}	\\
$	2454019.33015	$	&	$	0.00040	$	&	$	4522.5	$	&	$	-0.0125	$	&	\cite{2007IBVS.5781....1D}	\\
$	2454225.93715	$	&	$	0.00030	$	&	$	5137.5	$	&	$	-0.0147	$	&	\cite{2008IBVS.5820....1N}	\\
$	2457510.83979	$	&	$	0.00020	$	&	$	14915.5	$	&	$	-0.0312	$	&	\cite{2017IBVS.6195....1N}	\\
$	2457890.79250	$	&	$	0.00020	$	&	$	16046.5	$	&	$	-0.0379	$	&	\cite{2018IBVS.6234....1N}	\\
$	2457996.45310	$	&				&	$	16361	$	&	$	-0.0336	$	&	B.R.N.O.	\\
$	2459070.29269	$	&	$	0.00324	$	&	$	19557.5	$	&	$	-0.0582	$	&	This study	\\
$	2459070.46574	$	&	$	0.00030	$	&	$	19558	$	&	$	-0.0531	$	&	This study	\\
$	2459076.34273	$	&	$	0.00034	$	&	$	19575.5	$	&	$	-0.0552	$	&	This study	\\
$	2459076.50843	$	&	$	0.00030	$	&	$	19576	$	&	$	-0.0575	$	&	This study	\\
\hline
\hline
\end{tabular}
\end{center}
\label{A6}
\end{table}

\clearpage
\bibliographystyle{aasjournal}
\bibliography{new.ms}

\end{document}